\documentclass[12pt]{article}
\usepackage{epsfig,amsmath,amsfonts,amssymb,amstext,afterpage,psfrag,
}
\usepackage{slashed}
\usepackage[square,comma,sort&compress,numbers]{natbib}
\newcommand{\myv}[1]{{{\rule{0cm}{0.55cm}#1}}}
\allowdisplaybreaks
\usepackage{hyperref}
\newcommand{\vp}{\varphi}
\newcommand{\hs}{\vphantom{\dfrac{1}{2}}}

\newcommand{\wt}{\widetilde}

\newdimen\TW
\usepackage{color}
\definecolor{light-gray}{gray}{0.9}
\definecolor{dark-gray}{gray}{0.7}
\usepackage{tikz}

\long\def\symbolfootnote[#1]#2{\begingroup%
\def\thefootnote{\fnsymbol{footnote}}\footnote[#1]{#2}\endgroup}

\def\spose#1{\hbox to 0pt{#1\hss}}
\def\lsim{\mathrel{\spose{\lower 3pt\hbox{$\mathchar"218$}}
 \raise 2.0pt\hbox{$\mathchar"13C$}}}
\def\gsim{\mathrel{\spose{\lower 3pt\hbox{$\mathchar"218$}}
 \raise 2.0pt\hbox{$\mathchar"13E$}}}

\begin{document}

\begin{titlepage}

\begin{flushright}
{\small
LMU-ASC~15/19\\ 
FERMILAB-PUB-19-003-T\\
April 2019
}
\end{flushright}

\vspace{0.5cm}
\begin{center}
{\Large\bf \boldmath                                               
Master formula for one-loop renormalization\\
[0.15cm] of bosonic SMEFT operators
\unboldmath}
\end{center}

\vspace{0.5cm}
\begin{center}
{Gerhard Buchalla$^a$, Alejandro Celis$^{a}$, Claudius Krause$^b$ and 
Jan-Niklas Toelstede$^{a,c}$} 
\end{center}

\vspace*{0.4cm}

\begin{center}
$^a$Ludwig-Maximilians-Universit\"at M\"unchen, Fakult\"at f\"ur Physik,\\
Arnold Sommerfeld Center for Theoretical Physics, 
D--80333 M\"unchen, Germany\\
\vspace*{0.2cm}
$^b$Theoretical Physics Department, Fermi National Accelerator Laboratory,\\ 
Batavia, IL, 60510, USA\\
\vspace*{0.2cm}
$^c$Physik Department T31, James-Franck-Stra\ss e 1,\\ 
Technische Universit\"at M\"unchen,\\
D--85748 Garching, Germany
\end{center}

\vspace{1.5cm}
\begin{abstract}
\vspace{0.2cm}\noindent

Using background-field method and super-heat-kernel expansion, we derive a 
master formula for the one-loop UV divergences of the bosonic dimension-6 
operators in Standard Model Effective Field Theory (SMEFT).  This approach 
reduces the calculation of all the UV divergences to algebraic manipulations.  
Using this formula we corroborate results in the literature for the one-loop 
anomalous dimension matrix of SMEFT obtained via diagrammatic methods, 
considering contributions from the operators 
$X^3, \phi^6, \phi^4 D^2, X^2 \phi^2$ of the Warsaw basis. The formula is 
derived in a general way and can be applied to other quantum field theories
as well.

\end{abstract}

\vfill

\end{titlepage}

\section{Introduction}
\label{sec:intro}

In the background field method, the one-loop effective action for a general 
quantum field theory involving fermions and bosons can be written elegantly 
in terms of the super-determinant of a fluctuation 
operator~\cite{Jack:1984vj,Lee:1984ud,Neufeld:1998js}.  This formulation is 
particularly suitable for the calculation of the ultraviolet (UV) one-loop 
divergences of the theory, as the problem is reduced to algebraic manipulations
in which gauge invariance for the background fields is manifest during all 
stages of the computation~\cite{Abbott:1981ke}.   These methods have been 
used in the one-loop renormalization of effective theories for QCD 
at low energies~\cite{Neufeld:1998mb,Knecht:1999ag} and, more recently, 
for effective theories of the Higgs sector~\cite{Buchalla:2017jlu}.       

When the fluctuation operator can be cast in the minimal form 
$D_{\mu} D^{\mu} + Y$, the one-loop divergences are expressed with a compact 
formula (in terms of the standard expression for the second Seeley--DeWitt
 coefficient~\cite{Jack:1984vj,Lee:1984ud,Neufeld:1998js}).   Renormalizable 
gauge theories fall in this category if an appropriate choice is made for the 
gauge fixing of the fluctuating fields~\cite{tHooft:1973bhk}.   When the 
fluctuation operator is non-minimal, obtaining a closed expression for the 
one-loop divergences is a much more challenging task.  It is not hard to 
find these situations. For instance, gauge theories with a general 
$R_{\xi}$-gauge fixing for the fluctuating fields will give non-minimal 
operators. A calculation of this type has been done within the context of 
chiral perturbation theory in~\cite{Agadjanov:2013lra}.



In this work we are interested in the one-loop renormalization of the 
Standard Model Effective Field Theory  (SMEFT) at the level of dimension-6 
operators~\cite{Buchmuller:1985jz,Grzadkowski:2010es}.    The complete 
derivation of the one-loop anomalous dimension matrix in this case was 
accomplished via diagrammatic methods 
in~\cite{Jenkins:2013zja,Jenkins:2013wua,Alonso:2013hga,Alonso:2014zka}.   
We present here a first step in the program of performing this calculation
relying on functional methods.       Using background field method and 
super-heat-kernel expansion, we derive a master formula for the one-loop 
UV divergences of all the operators of the type 
$X^3, \phi^6, \phi^4 D^2, X^2 \phi^2$ from the Warsaw 
basis~\cite{Grzadkowski:2010es}.   Some of these operators have multiple 
derivatives and introduce non-minimal structures in the bosonic part of the 
fluctuation operator.        We will present the master formula in general 
terms as it can be useful within other quantum field theories.    
The results presented in this article have also been discussed in the 
thesis~\cite{Janmaster}.

Using the master formula we provide a non-trivial cross-check of the results 
in~\cite{Jenkins:2013zja,Jenkins:2013wua,Alonso:2013hga}, finding agreement 
with the renormalization group equations (RGEs) when taking into account 
contributions from the operators $X^3, \phi^6, \phi^4 D^2, X^2 \phi^2$. 
Such cross-checks are relevant given the phenomenological impact of 
renormalization group effects when analyzing scenarios of physics beyond the 
Standard Model (SM)~\cite{Grojean:2013kd,Alonso:2013hga,Celis:2017hod}. 
 
The article is organized as follows. In Sec.~\ref{sec:mast} we review 
the super-heat-kernel approach to the one-loop effective action 
and derive a master formula for the UV divergences.    
In Sec.~\ref{sec:SMEFT} we provide the context for the application 
of the master formula to SMEFT. The derivation of the RGEs from the 
one-loop divergences is discussed in Sec.~\ref{sec:rge}.
In Sec.~\ref{sec:bosonic} we give details on our calculation 
of divergences and RGEs for the bosonic dimension-6 operators of SMEFT.
We summarize the results in Sec.~\ref{sec:summary} and 
conclude in Sec.~\ref{sec:conc}.    
App.~\ref{sec:app2} contains details on the real representation of the Higgs 
field. App.~\ref{sec:appSM} summarizes the building blocks of the master 
formula and the beta functions of the SM at 
dimension 4.

\section{Master formula} 
\label{sec:mast}
 
\subsection{One-loop effective action}

We consider a quantum field theory with real boson fields ($\phi$) and 
spin-$1/2$ fermion fields ($\psi$).    In the background field method, 
fields are expanded in terms of a background field that satisfies the classical equations of motion and a fluctuating field: $\phi \to \phi + \xi$,  
$\psi \to \psi  +  \eta$, 
$\bar \psi \to  \bar \psi +  \bar \eta$~\cite{Abbott:1981ke}.       
We assume that the action of the theory expanded to second order in the 
fluctuations can be written as
\begin{align}  \label{fluclag}
S^{(2)}=  \int d^{d} x \left[ -\frac{1}{2}\xi^TA\xi + \bar{\eta}B\eta + 
\bar{\eta}\Gamma\xi + \xi^T \bar{\Gamma} \eta  \right] \; .
\end{align} 
We work in $d$-dimensional Minkowski space with $d=4-2\epsilon$, 
following the usual prescription in dimensional regularization.   
The differential operators $A, B$ and the fermionic functions 
$\Gamma, \bar \Gamma$ depend on the background fields.    
One-loop corrections to the effective action are given 
by~\cite{Jack:1984vj,Lee:1984ud,Neufeld:1998js}
\begin{align}
\Gamma^{1\text{L}}_\text{eff}[\phi,\psi,\bar{\psi}]= \frac{i}{2}\text{Str ln }K\; 
\label{super_eff} \,,
\end{align}
with  
\begin{align}
K = \begin{pmatrix} A&\bar{\Gamma}&-\Gamma^T\\-\bar{\Gamma}^T&0&B^T\\\Gamma&-B&0 \end{pmatrix} \; . \label{supermatrix}
\end{align}
Here $\mathrm{Str}$ stands for supertrace. For a review of supermatrix algebra see~\cite{VanNieuwenhuizen:1981ae}.   We will distinguish supertraces with and without integration over Minkowski space $\mathrm{Str} \, O = \int d^{d} x \,   \mathrm{str}  \,  \langle x |  O |  x \rangle$.   It is useful to write the one-loop effective action in the form~\cite{Neufeld:1998js} 
\begin{align}  \label{eqtrick}
\Gamma^{1\text{L}}_\text{eff}[\phi,\psi,\bar{\psi}]   \equiv   \frac{i}{2}  \text{Str ln }     \Delta   =   \frac{i}{2}  \text{Str ln }   \begin{pmatrix} A&      \sqrt{2}   i \bar \Gamma   \gamma_5  B \gamma_5         \\   \sqrt{2} i \Gamma  &    B \gamma_5 B \gamma_5   \end{pmatrix} \, ,
\end{align}
which applies in the present case, where operators are at most bilinear
in fermion fields. 
When the differential operator takes the form 
$A =   (\partial^{\mu}  + N^{\mu})(\partial_{\mu}  + N_{\mu})   + M$ and 
$B = i \slashed{\partial} - F$, it is possible to cast the fluctuation 
operator $\Delta$ into the minimal form
\begin{align}  \label{minform}
\Delta = D^{\mu} D_{\mu}  + Y \,,
\end{align}
with $D_{\mu} =  \partial_{\mu} +  X_{\mu}$  
\begin{align}  \label{minform2}
X_{\mu}  &=     \begin{pmatrix}    N_{\mu}  &     \frac{1}{\sqrt{2}}   \bar \Gamma \gamma_{\mu}            \\     0  &    \frac{i}{2}  (    F \gamma_{\mu}   -  \gamma_{\mu}  F_5  )  \end{pmatrix}  \,,  \nonumber \\
Y  &= \begin{pmatrix} M & 
-\frac{1}{\sqrt{2}}\bar{\Gamma}
(\overset{\text{\footnotesize{$\leftarrow$}}}{\slashed{\partial}}
-\slashed{N}+\frac{i}{2} \gamma_\mu F \gamma^\mu )\\ 
\sqrt{2}i\Gamma & 
\frac{1}{4}(F\gamma_\mu F \gamma^\mu + \gamma_\mu F_5 \gamma^\mu F_5  - 
\gamma_\mu F_5 F\gamma^\mu) -\frac{i}{2}\partial_\mu(F\gamma^\mu +\gamma^\mu F_5) 
\end{pmatrix}\,.
\end{align}
Here we have defined $F_5 = \gamma_5 F \gamma_5$.     Furthermore, we will assume that $F$ can be decomposed as
 \begin{align}  \label{Fterm}
F = rP_R + lP_L +  R_\mu \gamma^\mu P_R + L_\mu \gamma^\mu P_L \; ,
\end{align}
with $P_{L,R}$ the chiral projectors.

\subsection{Minimal operator}

For the operator~\eqref{minform}, the one-loop divergences at the level of 
the Lagrangian $(\Gamma^{1\text{L}}_\text{eff} = \int d^dx~\mathcal{L}^{1})$ are 
given by the usual formula~\cite{Jack:1984vj,Lee:1984ud,Neufeld:1998js}
\begin{align}
\mathcal{L}^{1}_{\text{div}}(\Delta) = 
\frac{1}{16\pi^2(4-d)}\text{str}\left(\frac{1}{12} 
X_{\mu \nu} X^{\mu \nu} + \frac{1}{2} Y^2\right) \; , \label{one_loop_div_minimal}
\end{align}
with $X_{\mu\nu} = [D_{\mu},D_{\nu}]=  
\partial_{\mu} X_{\nu}-\partial_{\nu}X_{\mu}+[X_{\mu},X_{\nu}]$.   
Eq.~\eqref{one_loop_div_minimal} can be developed further by evaluating the 
supertrace and by performing the trace over Dirac indices. This leads 
to the well known 't Hooft formula~\cite{tHooft:1973bhk}
\begin{align}   
\mathcal{L}^{1}_{\text{div}}(\Delta) 
&= \frac{1}{ 16\pi^2 (4-d)} \text{tr}\bigg(\frac{1}{12}N_{\mu\nu}N^{\mu\nu} +\frac{1}{2}M^2 + i\bar{\Gamma}\slashed{D}\Gamma \nonumber \\&\qquad \qquad \qquad \qquad  +2(D_\mu l)(D^\mu r)-2(lr)^2 -\frac{1}{3}R_{\mu\nu}R^{\mu\nu} -\frac{1}{3}L_{\mu\nu}L^{\mu\nu} \bigg) \; . \label{master_formula}
\end{align}
Here we use the definitions
\begin{align}\begin{aligned}
D_\mu \Gamma &\equiv \partial_\mu \Gamma+N_\mu \Gamma-\frac{i}{2}F\gamma_\mu \Gamma \; , && N_{\mu\nu}\equiv\partial_\mu N_\nu - \partial_\nu N_\mu +[N_\mu,N_\nu] \; , \\ 
D_\mu r &\equiv \partial_\mu r +iL_\mu r-irR_\mu \; , \vphantom{\frac{i}{2}} &&R_{\mu\nu} \equiv \partial_\mu R_\nu - \partial_\nu R_\mu +i[R_\mu,R_\nu] \; ,\\
D_\mu l &\equiv \partial_\mu l +iR_\mu l-ilL_\mu \; , \vphantom{\frac{i}{2}} &&L_{\mu\nu} \equiv \partial_\mu L_\nu - \partial_\nu L_\mu +i[L_\mu,L_\nu] \; . \label{blocks_definition} \end{aligned}
\end{align}

\subsection{Extension for non-minimal operator}

We will consider a fluctuation operator that receives the following 
corrections in the bosonic block of the supermatrix 
\begin{align}\begin{aligned}    \label{diff_operator_gen1}
\Delta' &= \Delta + \begin{pmatrix}c_B + 2b_B^\mu D_\mu + a_B^{\mu\nu} D_\mu D_\nu & 0 \\ 0&0 \end{pmatrix} \; .
\end{aligned}
\end{align}
The operator $\Delta$ is assumed to be written in the minimal form \eqref{minform}.    Here $D_{\mu}  =  \partial_{\mu}  + N_{\mu}$ is the covariant derivative 
in the bosonic sector, $c_B$ and $a_B^{\mu \nu}=a_B^{\nu \mu}$ are symmetric in the field indices while $b_{B}^{\mu}$ is antisymmetric.  The terms $c_B$ and $b_B^\mu$ can be absorbed into $M$ and $N^{\mu}$ of $\Delta$, however it will be useful to keep these corrections separate.     The correction $a_B^{\mu\nu} D_\mu D_\nu$ introduces a non-minimal structure in $\Delta^{\prime}$, similar to what occurs in a gauge theory with a generic $R_{\xi}$ gauge fixing term.     
We treat the corrections ($c_B,b_B^{\mu},a_B^{\mu\nu}$) as first-order 
perturbations of $\Delta$, which is sufficient for the one-loop renormalization
of SMEFT at the level of dimension-6 operators.
This feature simplifies considerably the calculation.

In order to derive the one-loop UV divergences using the super-heat-kernel method, it is convenient to write $\Delta^{\prime}$ in the form
\begin{align}  \label{diff_operator_gen2}
\Delta'= \Delta + c+ 2b^\mu D_\mu + a^{\mu\nu}D_\mu D_\nu \,,
\end{align} 
where now the covariant derivates are acting in superspace $D_{\mu} =  \partial_{\mu} +  X_{\mu}$, with $X_{\mu}$ defined in \eqref{minform2}.   
The new terms ($c,b^{\mu},a^{\mu\nu}$) are given by
\begin{align} \begin{aligned}
c &= \begin{pmatrix}c_B & \sqrt{2}\bar{\Gamma} \slashed{b}_B -\frac{1}{\sqrt{2}} (\partial_\mu\bar{\Gamma}) a_B^{\mu\nu} \gamma_\nu + \frac{1}{\sqrt{2}}\bar{\Gamma} N_\mu a_B^{\mu\nu} \gamma_\nu +\frac{1}{\sqrt{2}}\bar{\Gamma}a_B^{\mu\nu} \gamma_\mu X_{22\nu} \\ 0& 0 \end{pmatrix} \; , \\
b^\mu & = \begin{pmatrix}b_B^\mu & -\frac{1}{\sqrt{2}} \bar{\Gamma}a_B^{\mu\nu}\gamma_\nu \\ 0& 0 \end{pmatrix} \; ,\\
a^{\mu\nu} &= \begin{pmatrix} a_B^{\mu\nu} & 0 \\ 0& 0 \end{pmatrix} \; . \end{aligned}
\end{align}
Here $a$, $b$ and $c$ are supermatrices and $X_{22\nu}$ is the lower right
entry of $X_\nu$.
 
The one-loop UV divergences associated with $c$ and $b^{\mu}$ can be obtained 
from Eq.~\eqref{one_loop_div_minimal} after a redefinition of $Y$ and $X_{\mu}$, 
keeping terms linear in $c$ and $b^{\mu}$.  We focus in the following on the 
corrections due to $a^{\mu \nu}$. To evaluate the divergences from $a^{\mu\nu}$ 
we use the integral representation (dropping an irrelevant constant)
\begin{align}  
\text{Str ln }\Delta'
=  -  \int d^dx \text{ str } \int_0^\infty \frac{d\tau}{\tau}
\langle x|e^{-\tau\Delta'}|x\rangle   \,, \; \label{heat_kernel}
\end{align}
and implement the heat-kernel expansion \cite{Donoghue:1992dd,Ball:1988xg}
\begin{align}
\langle x|e^{-\tau\Delta'}|x\rangle \equiv \frac{i}{(4\pi\tau)^\frac{d}{2}}
\sum_{n=0}^{\infty} a_n(x,x)  \tau^n \; . \label{heat_exp}
\end{align}
The one-loop UV divergences are contained in the hermitean part of the
second Seeley--DeWitt coefficient \cite{Ball:1988xg}
\begin{align}
\mathcal{L}^{1}_{\text{div}}(\Delta^{\prime}) =   \frac{1}{2  (4\pi)^2   (4-d)  } 
\mathrm{str} \,  ( a_2(x,x)   +  a_2(x,x)^{\dag}) \,.
\end{align} 
The latter can be obtained from \eqref{heat_kernel} by going to momentum space 
and extracting the coefficient of $\tau^2$:
\begin{align}\begin{aligned}
a_2(x,x) &= -i \int \frac{d^dp}{\pi^\frac{d}{2}} e^{p^2} e^{-\left(\tau \Delta + 
\sqrt{\tau} 2i p^\mu D_\mu +\tau a^{\mu\nu} D_\mu D_\nu + \sqrt{\tau}a^{\mu\nu} 
2i p_\mu D_\nu - a^{\mu\nu}p_\mu p_\nu \right)} \bigg|_{\tau^2} \; .
\end{aligned}
\end{align}
Rearranging terms in the final result, and including the contributions from
$b$ and $c$, we get the following one-loop UV divergences from $\Delta'$ 
\begin{align}\begin{aligned} 
\mathcal{L}_\text{div}^1(\Delta^{\prime}) =&\; \mathcal{L}_\text{div}^1(\Delta)  + \frac{1}{2} \frac{1}{(4\pi)^2(4-d) }  \text{str} \,\bigg(   cY + \frac{1}{3}X_{\mu\nu}[D^\mu,b^\nu] - Y[D_\mu, b^\mu] \\&-\frac{1}{24}a^\lambda_{~\lambda} X_{\mu\nu}X^{\mu\nu} +\frac{1}{6} a^{\mu\nu} X_{\mu\lambda}X_\nu^{\ \lambda} +\frac{1}{6}X_{\mu\lambda}[D_\nu,[D^\lambda,a^{\mu\nu}]]     \\
&-\frac{1}{4}a^\lambda_{~\lambda} Y^2 + \frac{1}{3}Y[D_\mu,[D_\nu,a^{\mu\nu}]] -\frac{1}{12}Y[D_\mu,[D^\mu,a^\lambda_{~\lambda} ]]   + \mathrm{h.c.} \bigg)   \;. \label{DSix_Divergent_Lagrangian} \end{aligned}
\end{align} 
The term $\mathcal{L}_\text{div}^1(\Delta)$ represents the divergences 
associated with the minimal operator $\Delta$ in \eqref{diff_operator_gen2}, 
these are given by \eqref{one_loop_div_minimal}.  The remaining terms contain 
the corrections introduced by $(c_B,b_B^{\mu},a_{B}^{\mu\nu})$.

We can simplify the result by evaluating the supertrace and arrive at a master 
formula for 
$\delta \mathcal{L}_\text{div}^1 \equiv  \mathcal{L}_\text{div}^1(\Delta^{\prime}) 
-  \mathcal{L}_\text{div}^1(\Delta)$:\\
\begin{align}
\begin{aligned}   
\delta \mathcal{L}_\text{div}^1 =    \frac{1}{(4\pi)^2(4-d)} 
&\text{tr}\bigg(c_B M +\frac{1}{3}N_{\mu\nu}[D^\mu,b_B^\nu] 
+i\bar{\Gamma} \slashed{b}_B \Gamma  
-\frac{1}{6}\bar{\Gamma}
i\overset{\text{\footnotesize{$\leftrightarrow$}}}{\mathcal{D}}_{a_B}\Gamma \\
& -\frac{1}{24} a^\lambda_{B \lambda} N_{\mu\nu} N^{\mu\nu} 
+\frac{1}{6} a_B^{\mu\nu} N_{\mu\lambda} N_\nu^{\ \lambda} 
+\frac{1}{6} N_{\mu\lambda} [D_\nu,[D^\lambda,a_B^{\mu\nu}]] \\
&-\frac{1}{4}a^\lambda_{B\lambda}M^2 -\frac{1}{12}M[D_\mu,[D^\mu,a^\lambda_{B\lambda}]] 
+ \frac{1}{3}M[D_\mu,[D_\nu, a_B^{\mu\nu}]] \bigg) \;.  
\label{bosonic_master_formula}
\end{aligned}
\end{align}
Here we use 
$N_{\mu\nu} =  \partial_{\mu} N_{\nu}-\partial_{\nu}N_{\mu}+[N_{\mu},N_{\nu}]$ and
\begin{align} \begin{aligned}
\bar{\Gamma} 
i\overset{\text{\footnotesize{$\leftrightarrow$}}}{\mathcal{D}}_{a_B} \Gamma 
&\equiv 
\bigg(i\bar{\Gamma}a_B^{\mu\nu}\gamma^\lambda (\partial^\rho + N^\rho) \Gamma 
-i\bar{\Gamma} \gamma^\lambda 
(\overset{\text{\footnotesize{$\leftarrow$}}}{\partial^\rho} - N^\rho) 
a_B^{\mu\nu}\Gamma\bigg) (g_{\mu\nu}g_{\lambda\rho} - g_{\mu\lambda} g_{\nu\rho}) 
\\ &\hphantom{=}
+\frac{1}{2} \bar{\Gamma}a_B^{\mu\nu} \gamma^\lambda F \gamma^\rho \Gamma 
(g_{\mu\nu} g_{\lambda\rho}+2g_{\mu\lambda}g_{\nu\rho}) \; .
\end{aligned}
\end{align}
To derive \eqref{bosonic_master_formula} we assumed that $a_{B}^{\mu \nu}$ and 
$N^{\mu}$ have a block diagonal structure with respect to the
bosonic variables from spin-0 and spin-1 fields.

In the particular case $a_{B}^{\mu \nu} = a_B  g^{\mu \nu}$, the master formula \eqref{bosonic_master_formula} simplifies considerably
\begin{align}\begin{aligned}    \label{simplmf}
\delta \mathcal{L}_\text{div}^1   = \frac{1}{(4\pi)^2(4-d)} &\text{tr}\bigg(c_B M +\frac{1}{3}N_{\mu\nu}[D^\mu,b_B^\nu] +i\bar{\Gamma} \slashed{b}_B \Gamma \\ &\hphantom{\text{tr }}- a_B M^2 -\frac{1}{2}\bar{\Gamma}a_B i\slashed{D} \Gamma + \frac{1}{2} \bar{\Gamma} i\overset{\text{\footnotesize{$\leftarrow$}}}{\slashed{D}} a_B \Gamma \bigg) \; ,
\end{aligned}
\end{align}
with the covariant derivatives on fermions $\Gamma$, $\bar\Gamma$
\begin{align} 
i\slashed{D} =
i\slashed{\partial} + i\slashed{N} + \frac{1}{2} \gamma_\mu F \gamma^\mu \qquad 
\text{and} \qquad i\overset{\text{\footnotesize{$\leftarrow$}}}{\slashed{D}} = 
i\overset{\text{\footnotesize{$\leftarrow$}}}{\slashed{\partial}}-i\slashed{N} 
-\frac{1}{2}\gamma_\mu F \gamma^\mu \; .
\end{align}

\section{Application to SMEFT}  
\label{sec:SMEFT}

Assuming that the SM degrees of freedom are the only
ones present at energies below a certain high-energy scale $\Lambda
\gg M_W$ where new dynamics enters, we can parametrize physics at low 
energies ($E\ll \Lambda$) by the effective Lagrangian
\begin{equation} \label{eq:SMEFT}
\mathcal{L} = \mathcal{L}_{\rm SM} + \frac{1}{\Lambda} \sum_k C_k^{(5)} Q_k^{(5)}  
+ \frac{1}{\Lambda^2} \sum_k C_k^{(6)} Q_k^{(6)} 
+ \mathcal{O}\left( \frac{1}{\Lambda^3} \right) \, .
\end{equation}
Here $\mathcal{L}_{\rm SM}$ is the renormalizable SM
Lagrangian, $Q_k^{(5,6)}$ are higher-dimensional operators of
dimensions $5$ and $6$, $C_k^{(5,6)}$ are their Wilson
coefficients. At dimension 5 there is only the so-called Weinberg operator. 
A non-redundant basis of dimension-6 operators was 
defined in~\cite{Grzadkowski:2010es}.    

We denote the $\mathrm{SU}(3)$ and $\mathrm{SU}(2)$ generators by 
$T^A=\frac{1}{2}\lambda^A$ and $\tau^a=\frac{1}{2}\sigma^a$, where $\lambda^A$ 
and $\sigma^a$ are the Gell-Mann and Pauli matrices, respectively.  The Higgs 
field is parametrized in terms of 4 real components $\vp_i $:
\begin{align}
H \equiv \frac{1}{\sqrt{2}}\left(\widetilde{\phi},\phi\right)  =   
i\tau^i\vp_i \; .
\end{align}
Here $\phi$ is the Higgs doublet and $\widetilde{\phi} = i \sigma_2 \phi^*$ 
its conjugate, $i\in\{0,1,2,3\}$ and we defined $\tau^0=-\frac{i}{2}\mathbf{1}$.
Using the real representation for the Higgs field we can write the 
SM Lagrangian as
(our sign conventions for covariant derivatives and field strengths
coincide with \cite{Grzadkowski:2010es})
\begin{align}
\begin{aligned}
\mathcal{L}_\text{SM} =&-\frac{1}{4}G^A_{\mu\nu}G^{A\mu\nu}-\frac{1}{4}W^a_{\mu\nu}W^{a\mu\nu} -\frac{1}{4}B_{\mu\nu}B^{\mu\nu} \\ &+ \frac{1}{2}(D_\mu \vp)_i (D^\mu \vp)_i +\frac{m^2}{2}\vp_i\vp_i -\frac{\lambda}{8}(\vp_i\vp_i)^2 \\ &+ 
\bar{\psi}i\slashed{D}\psi    -  \left(  \bar{\psi}\sqrt{2}H\mathcal{Y} P_R\psi + \text{h.c.}  \right) \vphantom{\frac{1}{4}}\; . \label{SM_compact}
\end{aligned}
\end{align}
Fermions have been collected into $\psi=(u,d,\nu,e)^T$, where every component 
is a Dirac spinor.  The Yukawa matrices are grouped into $\mathcal{Y}=
\text{diag}(\mathcal{Y}_u,\mathcal{Y}_d,\mathcal{Y}_\nu,\mathcal{Y}_e)$.   
We will take $\mathcal{Y}_\nu=0$ in the following.
More details about the real representation of the Higgs field used here are 
provided in App.~\ref{sec:app2}.  
We expand the fields around their background component as
\begin{align}
\begin{aligned}
G^{A\mu} &\to G^{A\mu} +\alpha^{A\mu}\; , \qquad  & \vp_i &\to \vp_{i}+\xi_i\;, \\
W^{a\mu} &\to W^{a\mu} +\omega^{a\mu}\;, \qquad  & \psi &\to \psi+\eta\;, \\
B^\mu &\to B^\mu +\beta^\mu\;, \qquad & \bar{\psi} &\to \bar{\psi} + \bar{\eta} 
\;. \label{SM_background}
\end{aligned}
\end{align}
The operator $\Delta$ in \eqref{diff_operator_gen2} will contain the fluctuations of the SM Lagrangian while the corrections generated by the dimension-6 SMEFT operators will be encoded in ($c_B,b_{B}^{\mu},a_B^{\mu\nu}$).     We group the bosonic fluctuations into $(\alpha^{A\alpha},\,\omega^{a\lambda},\,\beta^\sigma,\, \xi_i)$.    To write the bosonic kinetic term as in~\eqref{fluclag} we redefine the gauge field fluctuations $(\alpha^{A\mu},\omega^{a\mu},\beta^\mu) \to i (\alpha^{A\mu},\omega^{a\mu},\beta^\mu)$.     The SM fluctuation operator written in minimal form is given in App.~\ref{sec:appSM}.      

The master formula~\eqref{bosonic_master_formula} can be applied to all the 
SMEFT operators of the type~\cite{Grzadkowski:2010es}
\begin{align}
X^3,\, \phi^6, \,\phi^4 D^2, \,X^2 \phi^2 \,.
\end{align}
See Table~\ref{pbsot} for a detailed list.   
\begin{table} 
\caption{\textit{Dimension-6 bosonic operators of the SMEFT from the 
Warsaw basis~\cite{Grzadkowski:2010es}. 
Here $\tilde X_{\mu\nu}\equiv 1/2\, \epsilon_{\mu\nu\rho\sigma}X^{\rho\sigma}$,
$\epsilon_{0123}=+1$. \label{pbsot}}}
\begin{center}
\begin{tabular}{|c|c||c|c|c|c|}
\hline  
\multicolumn{2}{|c||}{$X^3$} & \multicolumn{2}{|c|}{$X^2 \phi^2$}         \\
\hline
$Q_G$  & $f^{ABC} G_\mu^{A \nu} G_\nu^{B \rho} G_\rho^{C \mu}$ & $Q_{\phi G}$ & $\phi^\dagger \phi G_{\mu \nu}^A G^{A \mu \nu}$    \\[0.1cm]
$Q_{\widetilde G}$ & $f^{ABC} \widetilde G_\mu^{A \nu} G_\nu^{B \rho} G_\rho^{C \mu}$ & $Q_{\phi B}$ & $\phi^\dagger \phi B_{\mu \nu} B^{\mu \nu}$   \\[0.1cm]
$Q_W$ & $\epsilon^{abc} W_\mu^{a \nu} W_\nu^{b \rho} W_\rho^{c \mu}$ & $Q_{\phi W}$  & $\phi^\dagger \phi W_{\mu \nu}^a W^{a \mu \nu}$ \\[0.1cm]     \cline{5-6} 
$Q_{\widetilde W}$ & $\epsilon^{abc} \widetilde W_\mu^{a \nu} W_\nu^{b \rho} W_\rho^{c \mu}$ &  $Q_{\phi W B}$  & $\phi^\dagger \sigma^a \phi W_{\mu \nu}^a B^{\mu \nu}$  \\[0.1cm]
\cline{1-2}  
\multicolumn{2}{|c||}{$\myv{\phi^6}$} & $Q_{\phi \widetilde G}$   & $\phi^\dagger \phi \widetilde G_{\mu \nu}^A G^{A \mu \nu}$   \\[0.1cm]
\cline{1-2}
 $Q_{\phi}$ &$\hspace{-1cm}\myv{\left( \phi^\dagger \phi \right)^3}$ & $Q_{\phi \widetilde B}$ & $\phi^\dagger \phi \widetilde B_{\mu \nu} B^{\mu \nu}$  \\[0.1cm]
 \cline{1-2}
\multicolumn{2}{|c||}{~$\phi^4 D^2$} & $Q_{\phi \widetilde W}$ &  $\phi^\dagger \phi \widetilde W_{\mu \nu}^a W^{a \mu \nu}$  \\[0.1cm]
\cline{1-2}
$Q_{\phi \Box}$ & $\myv{\left( \phi^\dagger \phi \right) \Box \left( \phi^\dagger \phi \right)}$ & $Q_{\phi \widetilde W B}$ & $\phi^\dagger \sigma^a \phi \widetilde W_{\mu \nu}^a B^{\mu \nu}$  \\[0.1cm]
$Q_{\phi D}$ & $\left( \phi^\dagger D^\mu \phi \right)^\ast \left( \phi^\dagger D_\mu \phi \right)$ &  &  \\
\cline{1-4}  
\end{tabular}
\end{center}
\end{table}
These give rise to a fluctuation 
operator that can be cast in the form~\eqref{diff_operator_gen1}.      
Using~\eqref{bosonic_master_formula} we  calculate the one-loop UV divergences 
generated by these operators, and evaluate the corresponding contributions 
to the renormalization group equations.     

Under renormalization the operators in Table~\ref{pbsot} not only mix
among each other, but also into dimension-6 operators 
of the classes $\psi^2 \phi^3$, $\psi^2 X\phi$ and $\psi^2\phi^2 D$,
in the notation of \cite{Grzadkowski:2010es}. Using the
representation $\psi$ for the SM fermions, introduced above, these
operators can be expressed in a compact form.
The operators in the class $\psi^2 \phi^3$ can be combined into a 
single term:
\begin{align}
(\phi^\dagger \phi) \bar{\psi}_L \sqrt{2}H C_{\psi\phi} \psi_R + \text{h.c.} \; ,
\end{align}
where, as in the case of the Yukawa matrices, the Wilson coefficients have 
been collected into a matrix in flavour space as 
$C_{\psi\phi} = \text{diag}(C_{u\phi},C_{d\phi},0,C_{e\phi})$. The same scheme can be 
applied to the $\psi^2 X\phi$ operators. The resulting structures 
and flavour matrices are given by
\begin{align} \begin{aligned}
&\bar{\psi} \sigma^{\mu\nu} G^A_{\mu\nu} T^A \sqrt{2}H C_{\psi G} \psi \; , \qquad 
&C_{\psi G} &= \text{diag}\left(C_{uG}, C_{dG}, 0, 0 \right) \; , \hs \\
&\bar{\psi} \sigma^{\mu\nu} W^a_{\mu\nu} \sigma^a \sqrt{2}H C_{\psi W} \psi \; , 
&C_{\psi W} &= \text{diag}\left(C_{uW}, C_{dW}, 0 ,C_{eW}\right) \; , \hs \\  
&\bar{\psi} \sigma^{\mu\nu} B_{\mu\nu} \sqrt{2}H C_{\psi B} \psi \; , 
&C_{\psi B}& = \text{diag}\left(C_{uB}, C_{dB}, 0 , C_{eB} \right) \; . \hs
\end{aligned}
\end{align}
Finally, using the real representation of the covariant derivatives in 
appendix \ref{sec:app2}, we can reduce the operators of the remaining 
class $\psi^2 \phi^2 D$ to the form
\begin{align} \begin{aligned}
-&2i\bar{\psi}_R \gamma^\mu C_{\phi\psi 2} \vp( \tau^1 t^1_{R} + 
\tau^2 t_{R}^2)D_\mu\vp \psi_R \; , \qquad  
&C_{\phi\psi 2}& = \text{diag}(C_{\phi ud}, C_{\phi du}, 0, 0) \; , \hs \\ 
&2i(\vp t^3_{R} D_\mu \vp) \bar{\psi}_R \gamma^\mu C_{\phi\psi} \psi_R \; , 
&C_{\phi \psi}&= \text{diag}(C_{\phi u}, C_{\phi d}, 0 ,C_{\phi e}) \; , \hs \\ 
&2i(\vp t^3_{R} D_\mu \vp) \bar{\psi}_L\gamma^\mu C_{\phi\psi}^{(1)} \psi_L \; , 
&C^{(1)}_{\phi \psi}&= \text{diag}(C^{(1)}_{\phi q},C^{(1)}_{\phi l}) \; ,\hs \\
&2i(\vp t_{L}^aD_\mu \vp) \bar{\psi}_L\sigma^a\gamma^\mu C_{\phi \psi}^{(3)}\psi_L 
\; , &C^{(3)}_{\phi \psi}&= \text{diag}(C^{(3)}_{\phi q},C^{(3)}_{\phi l}) \; . \hs 
\end{aligned}
\end{align}
The coefficient matrices are chosen such that the corresponding operators are 
hermitian. For this, we had to generalize the first operator that contains 
$\phi ud$.

In the following sections we discuss the renormalization group equations 
for SMEFT and present the calculation of the one-loop UV divergences for the 
bosonic dimension-6 operators. We find agreement with the renormalization 
group equations presented 
in~\cite{Jenkins:2013zja,Jenkins:2013wua,Alonso:2013hga}.\footnote{We take 
into account the errata in \cite{err}.}

\section{Renormalization group equations}
\label{sec:rge}

We next present a concise derivation of the one-loop renormalization group 
equations of dimension-6 operators in SMEFT.
The results will allow us to convert the UV divergences of the one-loop
corrections into the beta functions for the operator coefficients.
 
The coefficients and operators of the dimension-6 Lagrangian in 
(\ref{eq:SMEFT}) are renormalized as
\begin{equation}\label{c0q0}
C^{(0)}_i Q^{(0)}_i = Z_{ij} C_j Z^{(i)}_F Q_i =
C_i Q_i +Q_i (Z^{(i)}_F Z_{ij}-\delta_{ij}) C_j \, .
\end{equation}
The factor $Z^{(i)}_F$ collects the field-renormalization constants
coming from the fields of the operator $Q_i$. A summation over
$i$ and $j$ is understood. It will be convenient to define 
\begin{equation}\label{zzkk}
Z_{ij}=\delta_{ij} +\frac{K_{ij}}{32\pi^2\epsilon}\, ,\qquad\quad
Z^{(i)}_F = 1 + \frac{K_{(i)}}{32\pi^2\epsilon}
\end{equation}
for the one-loop $Z$ factors in minimal subtraction. Eq. (\ref{c0q0})
then implies for the counterterm Lagrangian
\begin{equation}\label{lct}
32\pi^2\epsilon\ {\cal L}_{\rm CT} =
Q_i\left( K_{(i)}\delta_{ij} + K_{ij}\right)\, C_j\,\frac{1}{\Lambda^2}
\equiv -32\pi^2\epsilon\ \delta{\cal L}^1_{\rm div} \, ,
\end{equation}
which is identical, up to a sign, to the divergent part of the one-loop 
effective Lagrangian $\delta{\cal L}^1_{\rm div}$, to be computed with
the methods discussed in section \ref{sec:mast}.

In order to relate the coefficients $K_{ij}$ in (\ref{lct}) to the
renormalization-group evolution of the $C_j$, we note that
\begin{equation}\label{c0zc}
C^{(0)}_i =\mu^{n_{(i)}\epsilon} Z_{ij} C_j \, .
\end{equation}
Here the scale $\mu$ makes explicit the mass dimension $n_{(i)}\epsilon$
of the unrenormalized coefficient $C^{(0)}_i$. 
$Z_{ij}$ and $C_j$ are dimensionless.
Differentiating (\ref{c0zc}) with respect to $t\equiv\ln\mu$, using
(\ref{zzkk}) and working to one-loop accuracy, we find for the 
beta functions of the operator coefficients
\begin{equation}\label{betai1}
\beta_i\equiv 16\pi^2\frac{dC_i}{dt}=-\frac{1}{2\epsilon}\frac{dK_{ij}}{dt}C_j
+\frac{n_{(j)}-n_{(i)}}{2} K_{ij} C_j \, .
\end{equation}

This expression can be simplified\footnote{An equivalent
discussion in terms of NDA weights has been given 
in~\cite{Jenkins:2013sda}.} using the concept of 
{\it chiral dimensions}, 
which keeps track of the loop order of terms in a quantum field 
theory~\cite{Buchalla:2013eza}.\footnote{The chiral dimension is 0 for a boson 
field, and 1 for a weak coupling, a derivative, or a fermion bilinear.
The loop order $L$ of a term is related to its total chiral dimension $\chi$
through $\chi=2 L + 2$ \cite{Buchalla:2013eza}.} Consider an operator
containing a number of field-strength factors $X_{\mu\nu}$, scalar fields
$\phi$, fermions $\psi$ and derivatives $D$, schematically
\begin{equation}\label{qxphi}
Q=X^{N_X}_{\mu\nu} \phi^{N_\phi} D^{N_D} \psi^{N_F} \, .
\end{equation}
In $4-2\epsilon$ space-time dimensions this operator has 
{\it canonical dimension}
\begin{eqnarray}\label{dimq}
d(Q) &=& N_X (2-\epsilon)+N_\phi(1-\epsilon)+N_D + 
             N_F\left(\frac{3}{2}-\epsilon\right)\nonumber\\
&=& 2 N_X + N_\phi + N_D +\frac{3}{2} N_F - (N_X+N_\phi+N_F)\epsilon \, .
\end{eqnarray}
In 4 space-time dimensions the canonical dimension is
$d_0(Q)=2 N_X + N_\phi + N_D + 3/2 N_F$, constrained to be $d_0(Q)=6$ for 
the case at hand. Since the Lagrangian term $C^{(0)} Q^{(0)}/\Lambda^2$
has canonical dimension $4-2\epsilon$, it follows that
\begin{equation}\label{dimc0}
d(C^{(0)}) = (N_X+N_\phi+N_F-2)\epsilon
\end{equation}
for the coefficient $C^{(0)}$ of this operator. Using the 
{\it chiral dimension}
of $Q$ in (\ref{qxphi}),
\begin{equation}\label{chidimq}
\chi(Q)=N_X + N_D + \frac{1}{2} N_F \, ,
\end{equation}
we have $d(C^{(0)})=(d_0(Q)-\chi(Q)-2)\epsilon =(4-\chi(Q))\epsilon$
and therefore
\begin{equation}\label{nichi}
n_{(i)}=4-\chi_i
\end{equation}
for the parameter $n_{(i)}$ defined in (\ref{c0zc}). The canonical
dimension of the coefficient $C^{(0)}$ can thus be expressed through 
the chiral dimension $\chi_i$ of the corresponding dimension-6 operator. 

Inspecting (\ref{lct}), we note that the one-loop corrections of an operator
$Q_j$ yield the divergent terms
\begin{equation}\label{ldivqj}
\Delta {\cal L}_{\rm div}[Q_j]=-\frac{1}{32\pi^2\epsilon}\sum_i Q_i
\left( K_{(i)}\, \delta_{ij} + K_{ij}\right)\, \frac{C_j}{\Lambda^2}\, .
\end{equation}
The chiral dimension of this expression equals the chiral dimension of $Q_j$ 
increased by 2, since it results from the insertion of $Q_j$ into one-loop
diagrams. Therefore
\begin{equation}\label{chij2}
\chi_i + \chi(K_{(i)}\, \delta_{ij} + K_{ij}) = \chi_j + 2 \, .
\end{equation}
It follows that
\begin{equation}\label{chikij}
\chi(K_{ij})=\chi_j - \chi_i + 2 \, .
\end{equation}
For $i=j$ we also have $\chi(K_{(j)})=\chi(K_{jj})=2$.

Since, by definition, fields and derivatives belong to the operators $Q_i$,
the chiral dimension of $K_{ij}$ can only arise from weak couplings.
Eq. (\ref{chikij}) implies that the coefficient $K_{ij}$ is proportional to
$\chi_j-\chi_i+2$ weak couplings $\kappa$, which may be gauge or Yukawa
couplings, or $\sqrt{\lambda}$. Each of those fulfills
$d\kappa/d\ln t= -\epsilon\kappa+{\cal O}(1/16\pi^2)$.
We then conclude that
\begin{equation}\label{dkijdt}
\frac{d}{dt}K_{ij}=(\chi_i-\chi_j-2)\epsilon\, K_{ij}\, .
\end{equation}
Inserting (\ref{nichi}) and (\ref{dkijdt}) into (\ref{betai1}), we finally
obtain
\begin{equation}\label{betakc}
\beta_i=K_{ij} C_j \, .
\end{equation}
This means that the contributions to the beta function for coefficient 
$C_i$ from the insertions of operator $Q_j$ can be simply read off from the
term $K_{ij}$ in the counterterm Lagrangian.
For the mixing among coefficients of dimension-6 operators with $i\not= j$,
(\ref{betakc}) can be applied immediately.
In the case of $i=j$, (\ref{lct}) shows that $K_{(i)}$ has to be subtracted
from the coefficient of the divergence ($K_{(i)}+K_{ii}$) to obtain 
the beta function entry $K_{ii}$. 
The $K_{(i)}$ are determined by the renormalization constants of the fields
composing the operator $Q_i$. 
The renormalization factors needed for the bosonic operators are
\begin{eqnarray}\label{zfield}
Z_{fW} &=& 1+\frac{g^2}{32\pi^2\epsilon}
\left(\frac{44}{3}-\frac{2}{3}(N_c+1)f-\frac{1}{3}\right) =
1+\frac{g^2}{32\pi^2\epsilon}\frac{19}{3}\nonumber\\
Z_{fB} &=& 1-\frac{g'^2}{32\pi^2\epsilon}
\left(\left(\frac{22N_c}{27}+2\right)f +\frac{1}{3}\right) =
1-\frac{g'^2}{32\pi^2\epsilon}\frac{41}{3}\nonumber\\
Z_{fG} &=& 1+\frac{g_s^2}{32\pi^2\epsilon} \frac{22 N_c-4 N_f}{3} =
1+\frac{g_s^2}{32\pi^2\epsilon} 14\nonumber\\
Z_{f\phi} &=& 1+\frac{6g^2 + 2 g'^2 - 2 \gamma_\phi}{32\pi^2\epsilon}
\end{eqnarray}
where $N_c=3$, $N_f=6$ and $f=3$ denotes the number of colours,
quark flavours and fermion families, respectively, and
\begin{equation}\label{gammaphi}
\gamma_\phi=N_c\, 
{\rm tr}({\cal Y}^\dagger_u {\cal Y}_u + {\cal Y}^\dagger_d {\cal Y}_d)
+ {\rm tr}({\cal Y}^\dagger_e {\cal Y}_e) 
\equiv\langle {\cal Y}^\dagger {\cal Y} \rangle \, .
\end{equation}
Using (\ref{zfield}), we find the coefficients $K_{(i)}$ for the
bosonic operators in Table \ref{pbsot}:
\begin{eqnarray}
K_{(G)} &=& K_{(\tilde G)} = 21 g^2_s\, ,\qquad\qquad
K_{(W)} = K_{(\tilde W)} = \frac{19}{2} g^2\nonumber\\
K_{(\phi)} &=& 6(3g^2+g'^2-\gamma_\phi)\, ,\qquad
K_{(\phi\Box)} = K_{(\phi D)}= 4(3g^2+g'^2-\gamma_\phi)\nonumber 
\end{eqnarray}
\begin{eqnarray}\label{kki}
K_{(\phi G)} &=& K_{(\phi\tilde G)} =14 g^2_s + 6 g^2 + 2 g'^2 -2\gamma_\phi
\nonumber\\
K_{(\phi W)} &=& K_{(\phi\tilde W)} =\frac{37}{3} g^2 + 2 g'^2 -2\gamma_\phi
\nonumber\\
K_{(\phi B)} &=& K_{(\phi\tilde B)} =6 g^2 - \frac{35}{3} g'^2 -2\gamma_\phi
\nonumber\\
K_{(\phi WB)} &=& 
K_{(\phi\tilde WB)}=\frac{55}{6} g^2-\frac{29}{6} g'^2-2\gamma_\phi \, .
\end{eqnarray}

Due to the presence of the mass parameter $m$ in the leading-order
Lagrangian, a dimension-6 operator $Q_j\equiv Q^{(6)}_j$ can also mix
into dimension-4 operators $Q^{(4)}_i$. Such terms are generated from
the one-loop corrections to $Q_j$ in the form $m^2 Q^{(4)}_i$,
which may formally be viewed as a dimension-6 operator. The master formula
for the beta functions in (\ref{betakc}) also applies in this case,
once the normalization of the coupling associated with $Q^{(4)}_i$ has been
properly taken into account.
In particular, from the divergent one-loop corrections to $Q_j$ proportional
to dimension-4 terms,
\begin{equation}\label{ldiv4qj}
32\pi^2\epsilon\, \Delta {\cal L}^{(4)}_{\rm div}[Q_j] =
-\sum_i m^2 Q^{(4)}_i K_{ij} \frac{C_j}{\Lambda^2}\, ,
\end{equation}
we find the contribution to the beta function
\begin{equation}\label{betakc4}
\beta^{(4)}_i \supseteq \frac{m^2}{\Lambda^2} K_{ij} C_j
\end{equation}
for $i=m^2$, $\lambda$, ${\cal Y}_{rs}$, where
\begin{equation}\label{qqi4}
Q^{(4)}_{m^2} =\phi^\dagger\phi,\qquad 
Q^{(4)}_{\lambda} =-\frac{1}{2}(\phi^\dagger\phi)^2,\qquad
Q^{(4)}_{{\cal Y}_{rs}} = -\sqrt{2} \bar\psi^r_L H \psi^s_R \, .
\end{equation}

From the gauge-kinetic terms
\begin{equation}\label{qgxx}
Q^{(4)}_{g_X} = X^a_{\mu\nu} X^{a\mu\nu}
\end{equation}
the beta function of the corresponding gauge coupling $g_X$
receives the contribution
\begin{equation}\label{betagx}
\beta_{g_X} \supseteq 2 g_X\frac{m^2}{\Lambda^2} K_{g_X j} C_j \, .
\end{equation}

Alternatively, we can apply the equations of motion of the scalar to set 
$Z_{f\phi} = 1$. For consistency, these have to be applied at the 
dimension-6 level, generating additional contributions:
\begin{equation}
\frac{1}{2} D_\mu \varphi_i D_\mu \varphi^i (-6 g^2 - 2 g^{\prime 2}+ 2 \gamma_\phi)
\frac{1}{16\pi^2 \epsilon}\rightarrow \frac{1}{16\pi^2 \epsilon} 
(3 g^2 + g^{\prime 2}- \gamma_\phi) \frac{\delta}{\delta \varphi}
(\mathcal{L}_4 + \mathcal{L}_6 ),
\end{equation}
yielding the same beta functions as the previously discussed procedure 
in all cases.

\section{Bosonic operators}
\label{sec:bosonic}

\subsection{Renormalization of the operator class $X^3$}

We begin our computation\footnote{For cross-checks of our 
calculations, the programs FeynCalc \cite{Mertig:1990an,Shtabovenko:2016sxi}
and Mathematica \cite{wolframresearch} proved useful,
as well as the compilation of formulas in \cite{Borodulin:2017pwh}.}
by calculating the one-loop renormalization of the $X^3$-ope\-ra\-tors. 
The four operators of this class contain the field strengths 
of the gauge groups $SU(3)_C$ and $SU(2)_L$. To apply our algorithm, it is 
useful to distinguish operators with and without dual field strength. 
The group structure is the same in both cases, and we only have to work out 
the fluctuation matrices once. The divergences and the renormalization 
are then obtained for the two cases in an analogous way.
\subsubsection{$G$- and $W$-operators}
We start with the operator\footnote{Throughout this chapter, we will 
sometimes drop the distinction of upper and lower Lorentz indices
for notational convenience.}

\begin{align}
Q_G = f^{ABC} G_{\mu\nu}^A G_{\nu\lambda}^B G_{\lambda\mu}^C \; . \label{G_operator}
\end{align}
The symmetrized fluctuation matrices from (\ref{diff_operator_gen1}) 
are given by
\begin{align} \begin{aligned}
c_{(A\alpha)(B\beta)} =&\ 6g_s f^{ABC}f^{CDE} G_{\alpha\lambda}^D G_{\lambda\beta}^E + 6g_sf^{ADC}f^{BEC} G_{\alpha\lambda}^D G_{\lambda\beta}^E \hs \\ &+3g_s f^{ADC}f^{BEC} G^D_{\mu\nu} G^E_{\mu\nu} g_{\alpha\beta} \; , \hs \\
b^\lambda_{(A\alpha)(B\beta)} =&\ \frac{3}{2}f^{ABC} \left( D_\alpha G_{\beta\lambda} + D_\beta G_{\alpha\lambda} + g^{\alpha\lambda}D_\mu G_{\mu\beta} + g^{\beta\lambda}D_\mu G_{\mu\alpha}\right)^C \\ &-3f^{ABC} (D_\mu G^{\mu\lambda})^C g_{\alpha\beta} \; , \hs \\
a^{\mu\nu}_{(A\alpha)(B\beta)} =&\ 3f^{ABC}\left(g^{\mu\alpha} G_{\nu\beta} + g^{\nu\alpha}G_{\mu\beta} - g^{\mu\beta}G_{\nu\alpha} - g^{\nu\beta} G_{\mu\alpha} -2 G_{\alpha\beta} g^{\mu\nu}\right)^C \hs \; . \label{G_fluctuations}
\end{aligned}
\end{align}
Here and in the following, the matrices $a_B$, $b_B$, $c_B$ from
(\ref{diff_operator_gen1}) will simply be denoted as $a$, $b$, $c$.
Since we are concerned with purely bosonic operators,
no confusion can arise in the present context.
Only the non-zero entries of these matrices are quoted explicitly.

\subsubsection*{Divergences and RGEs from $G$}

We evaluate the constituents of the master 
formula~(\ref{bosonic_master_formula}) for the $G$-operator:
\begin{align}
\begin{aligned}
\text{tr }cM &= -18g_s^2 C_3^\text{ad} f^{ABC} G^A_{\mu\nu} G^B_{\nu\lambda}G^C_{\lambda\mu} \; , \hs \\
-\frac{1}{3}\text{tr }[D^\mu,N_{\mu\nu}]b^\nu &= -2g_s C_3^\text{ad} (D_\mu G^{\mu\lambda})^A(D^\nu G_{\nu\lambda})^A \; , \\ -\frac{1}{4}\text{tr }a^{\lambda}_{\ \lambda}M^2 &= 6g_s^2 C_3^\text{ad} f^{ABC} G^A_{\mu\nu} G^B_{\nu\lambda}G^C_{\lambda\mu} \; , \\ -\frac{1}{12}\text{tr }a^\lambda_{\ \lambda}[D_\mu,[D^\mu,M]] &= 2g_s C_3^\text{ad} G_{\mu\nu}^A(D_\lambda D^\lambda G^{\mu\nu})^A \; , \\ +\frac{1}{3}\text{tr }a^{\mu\nu} [D_\mu, [D_\nu,M]] &= 4g_s^2 C_3^\text{ad} f^{ABC} G^A_{\mu\nu} G^B_{\nu\lambda}G^C_{\lambda\mu} \; , \\ \text{tr }i\bar{\Gamma} \slashed{b} \Gamma &= 6g_s C_3^\text{ad} (D_\mu G^{\mu\lambda})^A(D^\nu G_{\nu\lambda})^A \hs \; , \\ -\frac{1}{6}\text{tr }\bar{\Gamma} i\overset{\text{\footnotesize{$\leftrightarrow$}}}{\mathcal{D}}_{a} \Gamma &= -9g_s^2\bar{q}_L \sigma^{\mu\nu}G^A_{\mu\nu}T^A \sqrt{2}H\mathcal{Y} q_R + \text{h.c.}
\end{aligned}
\end{align}
To obtain the first equality we used the identity 
$f^{ADE}f^{BEF}f^{CFD} = C_3^\text{ad}/2 f^{ABC}$ for the structure constants,
where $C_N^\text{ad}=N$. 
Building blocks that vanish during the calculation are not explicitly 
listed. Summing the terms and using the equations of motion, we find the 
divergent Lagrangian
\begin{align} \begin{aligned}
32\pi^2\epsilon \mathcal{L}_G^\text{div} = \frac{C_G}{\Lambda^2} 
\bigg(&-36g_s^2f^{ABC} G^A_{\mu\nu} G^B_{\nu\lambda}G^C_{\lambda\mu} \\ 
&-9g_s^2\bar{q}_L \sigma^{\mu\nu}G^A_{\mu\nu}T^A \sqrt{2}H\mathcal{Y} q_R + 
\text{h.c.} \bigg) \; . \end{aligned}
\end{align}
Using the results of Sec.~\ref{sec:rge}, the contributions to the 
beta functions induced by these divergences are given by
\begin{align}
\begin{aligned}
\beta_G \supseteq 15g_s^2 C_G \; , \qquad 
\beta_{q G} \supseteq 9g_s^2C_G \mathcal{Y}_q \; .
\end{aligned}
\end{align}

\subsubsection*{Divergences and RGEs from $W$}

The $W$-operator is 
\begin{align}
Q_W = \epsilon^{abc} W_{\mu\nu}^a W_{\nu\lambda}^b W^c_{\lambda\mu} \; .
\end{align}
This operator is analogous to (\ref{G_operator}), and we simply 
adapt the fluctuations (\ref{G_fluctuations}) to the case of $SU(2)$. 
For this operator, the terms in the master formula then become
\begin{align}\begin{aligned}
\text{tr }cM =& -18g^2 C_2^\text{ad} \epsilon^{abc} W_{\mu\nu}^a W_{\nu\lambda}^b W^c_{\lambda\mu} +3g^3 C_2^\text{ad} (\phi^\dagger \phi) W^a_{\mu\nu} W^{a\mu\nu} \; ,\hs \\
-\frac{1}{3}\text{tr }[D^\mu,N_{\mu\nu}]b^\lambda  =& -2g C_2^\text{ad} (D_\mu W^{\mu\lambda})^a(D^\nu W_{\nu\lambda})^a \; , \\ -\frac{1}{4} \text{tr }a^\lambda_{\ \lambda}M^2 =&\  6g^2C_2^\text{ad}\epsilon^{abc} W_{\mu\nu}^a W_{\nu\lambda}^b W^c_{\lambda\mu} +6g^3C_2^\text{ad} (\phi^\dagger \phi) W^a_{\mu\nu} W^{a\mu\nu}  \\ &-6ig^2 C_2^\text{ad} W^a_{\mu\nu} (D^\mu\vp) t_L^a (D^\nu \vp) \hs \; , \\ -\frac{1}{12}\text{tr }a^\lambda_{\ \lambda}[D_\mu,[D^\mu,M]] =&\ 2gC_2^\text{ad} W^a_{\mu\nu} (D_\lambda D^\lambda W^{\mu\nu})^a \; , \\ \frac{1}{3}\text{tr }a^{\mu\nu}[D_\mu,[D_\nu,M]] =&\ 4g^2 C_2^\text{ad} \epsilon^{abc} W_{\mu\nu}^a W_{\nu\lambda}^b W^c_{\lambda\mu}\; , \\ \text{tr }i\bar{\Gamma}\slashed{b}\Gamma =&\ 6gC_2^\text{ad}(D_\mu W^{\mu\lambda})^a(D^\nu W_{\nu\lambda})^a \hs \\ &-6ig C_2^\text{ad}(D_\mu W^{\mu\nu})^a (\vp t_L^a D_\nu\vp)  \; . \hs
\end{aligned}
\end{align}
In total, we find the divergent Lagrangian
\begin{align}
\begin{aligned}
32\pi^2\epsilon\mathcal{L}_W^\text{div} = \frac{C_W}{\Lambda^2}\bigg(&-24g^2\epsilon^{abc} W_{\mu\nu}^a W_{\nu\lambda}^b W^c_{\lambda\mu} +15g^3(\phi^\dagger\phi) W^a_{\mu\nu}W^{a\mu\nu} \\ &-3g^2g'(\phi^\dagger \sigma^a \phi) W^a_{\mu\nu} B^{\mu\nu} \bigg) \; .
\end{aligned}
\end{align}
This gives rise to the contributions
\begin{align}
\beta_W \supseteq \frac{29}{2}g^2 C_W \; , \qquad 
\beta_{\phi W} \supseteq -15g^3C_W  \;, \qquad 
\beta_{\phi W B} \supseteq 3g^2 g' C_W \; .
\end{align}

\subsubsection{$\wt{G}$- and $\wt{W}$-operators}
The operators with dual field strength read
\begin{align}
Q_{\wt{G}} = f^{ABC} \wt{G}^A_{\mu\nu} G^B_{\nu\lambda} G^C_{\lambda\mu} \; . 
\label{Gtilde_operator}
\end{align}
To work out the fluctuation Lagrangian, it is useful to relate the 
different kinds of tensors by
\begin{align}
\epsilon^{\alpha\beta\mu\nu} G^A_{\nu\lambda} = g^{\alpha\lambda} \wt{G}^A_{\mu\beta} - g^{\beta\lambda} \wt{G}^A_{\mu\alpha} -g^{\mu\lambda} \wt{G}^A_{\alpha\beta} \; .
\end{align}
We find fluctuation matrices similar to (\ref{G_fluctuations}), namely
\begin{align}\begin{aligned}
c_{(A\alpha)(B\beta)} =&\ 3g_sf^{ADC} f^{CBE} (\wt{G}^D_{\alpha\mu} G^E_{\mu\beta} + G^D_{\alpha\mu} \wt{G}^E_{\mu\beta} ) + 3g_sf^{ADC}f^{CBE} \wt{G}^D_{\mu\nu} G^{E\mu\nu}g_{\alpha\beta}  \hs \\ &+3f^{ABC}(D_\mu D^\mu \wt{G}_{\alpha\beta})^C \; , \hs \\
b^\lambda_{(A\alpha)(B\beta)}=&\ \frac{3}{2}f^{ABC} (D_\alpha \wt{G}_{\beta\lambda} + D_\beta \wt{G}_{\alpha\lambda})^C \; , \\ a^{\mu\nu}_{(A\alpha)(B\beta)} =&\ 3f^{ABC} ( g^{\mu\alpha} \wt{G}_{\nu\beta} + g^{\nu\alpha}\wt{G}_{\mu\beta} - g^{\mu\beta}\wt{G}_{\nu\alpha} - g^{\nu\beta} \wt{G}_{\mu\alpha} -2\wt{G}_{\alpha\beta} g^{\mu\nu})^C \; . \hs \label{Gtilde_fluctuations}
\end{aligned}
\end{align}
Note that several terms vanish due to the Bianchi identity, 
which implies $(D_\mu \wt{G}^{\mu\nu})^A=0$. The computation simplifies in
comparison with the $G$- and $W$-operators.

\subsubsection*{Divergences and RGEs from $\wt{G}$}

For (\ref{Gtilde_fluctuations}), the non-zero parts of the master formula are
\begin{align}\begin{aligned}
\text{tr }cM &= -18g_s^2 C_3^\text{ad} f^{ABC} \wt{G}^A_{\mu\nu} G^B_{\nu\lambda} G^C_{\lambda\mu}\;, \hs \\
-\frac{1}{4}\text{tr }a^\lambda_{\ \lambda}M^2 &= 6g_s^2 C_3^\text{ad} f^{ABC} \wt{G}^A_{\mu\nu} G^B_{\nu\lambda} G^C_{\lambda\mu} \; , \\ -\frac{1}{12} \text{tr }a^\lambda_{\ \lambda}[D_\mu,[D^\mu,M]] &= -4g_s^2 C_3^\text{ad} f^{ABC} \wt{G}^A_{\mu\nu} G^B_{\nu\lambda} G^C_{\lambda\mu} \; , \\ \frac{1}{3} \text{tr }a^{\mu\nu} [D_\mu,[D_\nu,M]] &= 4g_s^2 C_3^\text{ad}f^{ABC} \wt{G}^A_{\mu\nu} G^B_{\nu\lambda} G^C_{\lambda\mu} \; , \\ -\frac{1}{6}\text{tr }\bar{\Gamma} i\overset{\text{\footnotesize{$\leftrightarrow$}}}{\mathcal{D}}_{a} \Gamma &= -9ig_s^2 \bar{q}_L \sigma^{\mu\nu}G^A_{\mu\nu} T^A \sqrt{2} H\mathcal{Y} q_R + \text{h.c.}
\end{aligned}
\end{align}
We sum the terms and obtain 
\begin{align}
\begin{aligned}
32\pi^2\epsilon \mathcal{L}_{\wt{G}}^\text{div} = \frac{C_{\wt{G}}}{\Lambda^2} 
\bigg(&-36g_s^2f^{ABC} {\wt{G}}^A_{\mu\nu} G^B_{\nu\lambda}G^C_{\lambda\mu} \\ 
&-9ig_s^2\bar{q}_L \sigma^{\mu\nu}G^A_{\mu\nu}T^A \sqrt{2}H\mathcal{Y} q_R + 
\text{h.c.} \bigg) \; . \end{aligned}
\end{align}
In this case, the divergences result in 
\begin{align}
\beta_{\wt{G}}\supseteq 15g_s^2 C_{\wt{G}} \; , \qquad 
\beta_{q G} \supseteq 9ig_s^2 C_{\wt{G}}\mathcal{Y}_q \; .
\end{align}

\subsubsection*{Divergences and RGEs from $\wt{W}$}

The last operator of this class is
\begin{align}
Q_{\wt{W}} = \epsilon^{abc} \wt{W}^a_{\mu\nu} W^b_{\nu\lambda} W^c_{\lambda\mu} \; . 
\end{align}
Again, we translate the fluctuations from (\ref{Gtilde_fluctuations}) 
to $SU(2)$ and find the divergent pieces
\begin{align}
\begin{aligned}
\text{tr }cM =& -18g^2C_2^\text{ad} \epsilon^{abc} \wt{W}^a_{\mu\nu} W^b_{\nu\lambda} W^c_{\lambda\mu} +3g^3 C_2^\text{ad} (\phi^\dagger \phi) \wt{W}^a_{\mu\nu} W^{\mu\nu} \; , \hs \\ -\frac{1}{4}\text{tr }a^\lambda_{\ \lambda}M^2 =&\ 6g^2 C_2^\text{ad} \epsilon^{abc} \wt{W}^a_{\mu\nu} W^b_{\nu\lambda} W^c_{\lambda\mu} +\frac{9}{2}g^3C_2^\text{ad} (\phi^\dagger \phi) \wt{W}^a_{\mu\nu} W^{a\mu\nu} \; , \\ & -\frac{3}{2}g^2g' C_2^\text{ad} (\phi^\dagger \sigma^a \phi) \wt{W}^a_{\mu\nu} B^{\mu\nu} \; , \\ -\frac{1}{12}\text{tr }a^\lambda_{\ \lambda}[D_\mu,[D^\mu,M]] =& -4g^2C_2^\text{ad} \epsilon^{abc} \wt{W}^a_{\mu\nu} W^b_{\nu\lambda} W^c_{\lambda\mu} \; , \\ \frac{1}{3} \text{tr } a^{\mu\nu}[D_\mu,[D_\nu,M]] =&\ 4g^2C_2^\text{ad} \epsilon^{abc} \wt{W}^a_{\mu\nu} W^b_{\nu\lambda} W^c_{\lambda\mu} \; .
\end{aligned}
\end{align}
The terms combine to the divergent Lagrangian
\begin{align}
\begin{aligned}
32\pi^2\epsilon\mathcal{L}_{\wt{W}}^\text{div} = \frac{C_{\wt{W}}}{\Lambda^2}
\bigg(&-24g^2\epsilon^{abc} \wt{W}_{\mu\nu}^a W_{\nu\lambda}^b W^c_{\lambda\mu} +
15g^3(\phi^\dagger \phi) \wt{W}^a_{\mu\nu}W^{a\mu\nu} \\ 
&-3g^2g'(\phi^\dagger \sigma^a \phi) \wt{W}^a_{\mu\nu} B^{\mu\nu} \bigg) \; .
\end{aligned}
\end{align}
The divergences lead to the renormalization group contributions
\begin{align}
\beta_{\wt{W}} \supseteq \frac{29}{2}g^2 C_{\wt{W}} \; , \qquad 
\beta_{\phi \wt{W}} \supseteq -15g^3C_{\wt{W}}  \;, \qquad 
\beta_{\phi \wt{W} B} \supseteq 3g^2 g' C_{\wt{W}} \; .
\end{align}

\subsection{Renormalization of the operator class $X^2\phi^2$}

We next consider the operators of the class $X^2\phi^2$. 
We again divide the computation into several steps, in a similar way
as in the discussion of the class $X^3$.

\subsubsection{$\phi X$-operators}

We first consider the operator
\begin{align}
Q_{\phi G} = (\phi^\dagger \phi)G^A_{\mu\nu} G^{A\mu\nu} \; .
\end{align}
We work out the fluctuation matrices and find 
($\vp^2\equiv \varphi_i \varphi_i$)
\begin{align} \begin{aligned}
c_{(A\alpha)(B\beta)} &= -3g_sf^{ABC}G_{\alpha\beta}^C\vp^2  + \delta^{AB} g_{\alpha\beta} 
\Box(\vp^2)-\frac{1}{2}\delta^{AB} \{ \partial_\alpha,\partial_\beta \} (\vp^2) 
\; ,\\
c_{(A\alpha)i} &= 2i\vp_i (D^\mu G_{\mu\alpha})^A + 2i(D^\mu \vp)_i G^A_{\mu\alpha}
\; , \hs \\
c_{ij} &=-G^A_{\mu\nu} G^{A\mu\nu} \delta_{ij} \; , \hs \\
b^\mu_{(A\alpha)(B\beta)} &=\frac{1}{2}\delta^{AB} g_{\mu \alpha} \partial_\beta (\vp^2)
 - \frac{1}{2} \delta^{AB} g_{\mu \beta} \partial_\alpha (\vp^2) \; , \\
b^\mu_{(A\alpha)i} &= 2i G^{A}_{\mu\alpha} \vp_i \; , \hs\\
a^{\mu\nu}_{(A\alpha)(B\beta)} &= - \delta^{AB} S^{\alpha\beta\mu\nu}\vp^2  
\hs \label{phiG_fluc} \; .
\end{aligned}
\end{align}
In the last entry, we defined the tensor 
$S^{\alpha\beta\mu\nu} \equiv 2g^{\alpha\beta} g^{\mu\nu} - g^{\alpha\mu} g^{\beta\nu} - 
g^{\alpha\nu} g^{\beta\mu}$, which is symmetric under the exchange of 
$\alpha \leftrightarrow \beta$ and $\mu \leftrightarrow \nu$.

\subsubsection*{Divergences and RGEs from $\phi G$}

For the fluctuations above, the various terms of the master formula become
\begin{align} \begin{aligned}
\text{tr }cM =& - \left(6\lambda + 12g_s^2C_3^\text{ad} + \frac{3}{2}g^2 + 
\frac{1}{2}g'^2  \right) (\phi^\dagger \phi)  G^A_{\mu\nu} G^{A\mu\nu} \\ 
& +4m^2 G^A_{\mu\nu} G^{A\mu\nu} \; , \hs \\
-\frac{1}{4}\text{tr }a^\lambda_{\ \lambda} M^2 =&\ 12g_s^2 C_3^\text{ad} 
(\phi^\dagger\phi)G^A_{\mu\nu}G^{A\mu\nu} \; ,\\
-\frac{1}{24}\text{tr }a^\lambda_{\ \lambda}N_{\mu\nu} N^{\mu\nu} =& 
-2g_s^2 C_3^\text{ad} (\phi^\dagger\phi)G^A_{\mu\nu}G^{A\mu\nu} \; ,\\
\frac{1}{6}\text{tr }a^{\mu\nu} N_{\mu\lambda}N^{\nu\lambda} =&\ 2g_s^2 C_3^\text{ad} 
(\phi^\dagger\phi)G^A_{\mu\nu}G^{A\mu\nu} \; , \\
i\bar{\Gamma}\slashed{b}\Gamma =&\ 4g_s \bar{q}_L \sigma^{\mu\nu} 
G^A_{\mu\nu} T^A \sqrt{2} H \mathcal{Y} q_R + \text{h.c.}\; ,\hs \\
-\frac{1}{6}\text{tr }\bar{\Gamma} 
i\overset{\text{\footnotesize{$\leftrightarrow$}}}{\mathcal{D}}_{a} 
\Gamma  =& -24g_s^2 C_3^F (\phi^\dagger \phi) \bar{q}_L \sqrt{2}H\mathcal{Y} q_R 
+ \text{h.c.}
\end{aligned}
\end{align}
We observe that terms proportional to $C_3^\text{ad}$ cancel in the sum. 
From the remaining terms we obtain the divergent Lagrangian
\begin{align}\begin{aligned}
32\pi^2\epsilon\mathcal{L}_{\phi G}^\text{div} = \frac{C_{\phi G}}{\Lambda^2} 
\bigg(&-\left(6\lambda + \frac{3}{2}g^2 +\frac{1}{2}g'^2 \right)
(\phi^\dagger\phi)G^A_{\mu\nu} G^{A\mu\nu} + 4m^2G^A_{\mu\nu} G^{A\mu\nu} \\ 
&-32g_s^2(\phi^\dagger \phi)\bar{q}_L\sqrt{2}H\mathcal{Y} q_R +4g_s \bar{q}_L 
\sigma^{\mu\nu} G^A_{\mu\nu} T^A \sqrt{2}H\mathcal{Y}q_R + \text{h.c.} 
\hs \bigg)\; . \end{aligned}
\end{align}
The coefficient of operator $\phi G$ then contributes to the beta functions
\begin{align}\begin{aligned}
\beta_{\phi G} &\supseteq \left(6\lambda -14g_s^2 - \frac{9}{2}g^2 
-\frac{3}{2}g'^2 +2\gamma_\phi \right) C_{\phi G} \; , 
\qquad  &\beta_{q\phi} &\supseteq 32g_s^2\mathcal{Y}_q C_{\phi G} \hs \\
\beta_{g_s} &\supseteq - 8g_s \frac{m^2}{\Lambda^2} C_{\phi G}  \; , 
\qquad  &\beta_{q G} &\supseteq -4g_s\mathcal{Y}_q C_{\phi G}  \; .\hs 
\end{aligned}
\end{align}

\subsubsection*{Divergences and RGEs from $\phi W$}

In a similar way we treat the $\phi W$-operator 
\begin{align}
Q_{\phi W} = (\phi^\dagger \phi)W^a_{\mu\nu} W^{a\mu\nu} \; .
\end{align}
The fluctuation matrices can be adapted from (\ref{phiG_fluc}). 
In the case of $SU(2)$ they yield
\begin{align}
\begin{aligned}
\text{tr }cM =& - \left(6\lambda +\frac{7}{2}g^2+12g^2C_2^\text{ad} +\frac{1}{2}g'^2 \right) (\phi^\dagger \phi)W^a_{\mu\nu} W^{a\mu\nu} \\ & +4m^2 W^a_{\mu\nu} W^{a\mu\nu} -2gg'(\phi^\dagger\sigma^a \phi)W^a_{\mu\nu} B^{\mu\nu}  \\&+ 9g^2 (\phi^\dagger \phi) \Box (\phi^\dagger \phi) \; , \\ -\frac{1}{4}\text{tr }a^\lambda_{\ \lambda} M^2 =&\ 12g^2 C_2^\text{ad} (\phi^\dagger \phi)W^a_{\mu\nu} W^{a\mu\nu} + 18m^2g^2(\phi^\dagger \phi)^2 \hs\\ &  -\left(18\lambda g^2 -9g^4 - 3g^2g'^2\right) (\phi^\dagger \phi)^3 -9g^2(\phi^\dagger \phi) \Box (\phi^\dagger \phi)    \\ &-9g^2(\phi^\dagger \phi) \bar{\psi}_L\sqrt{2}H\mathcal{Y}\psi_R +\text{h.c.} \; , \\ -\frac{1}{24}\text{tr }a^\lambda_{\ \lambda}N_{\mu\nu} N^{\mu\nu} =&-2g^2 C_2^\text{ad}  (\phi^\dagger \phi)W^a_{\mu\nu} W^{a\mu\nu} \; , \\ \frac{1}{6}\text{tr }a^{\mu\nu} N_{\mu\lambda}N^{\nu\lambda} =&\ 2g^2 C_2^\text{ad} (\phi^\dagger \phi)W^a_{\mu\nu} W^{a\mu\nu}  \; ,\hs \\ -\frac{1}{12} \text{tr }a^\lambda_{\ \lambda}[D_\mu,[D^\mu,M]] =&\ 6g^2 (\phi^\dagger \phi)\Box(\phi^\dagger \phi) \hs \; , \\ \frac{1}{3}\text{tr }a^{\mu\nu}[D_\mu,[D_\nu,M]] =& -6g^2(\phi^\dagger \phi)\Box(\phi^\dagger \phi) \; , \hs \\ \text{tr }i\bar{\Gamma}\slashed{b}\Gamma =&\ g\bar{\psi}_L \sigma^{\mu\nu} W^a_{\mu\nu} \sigma^a \sqrt{2}H\mathcal{Y} \psi_R + \text{h.c.}
\end{aligned}
\end{align}
The divergent Lagrangian then reads
\begin{align}\begin{aligned}
32\pi^2\epsilon \mathcal{L}^\text{div}_{\phi W} = \frac{C_{\phi W}}{\Lambda^2} \bigg(&-\left(6\lambda + \frac{7}{2}g^2+\frac{1}{2}g'^2 \right)(\phi^\dagger \phi)W^a_{\mu\nu} W^{a\mu\nu} + 4m^2 W^a_{\mu\nu} W^{a\mu\nu}\\ &-2gg' (\phi^\dagger\sigma^a \phi) W^a_{\mu\nu} B^{\mu\nu} -\left(18\lambda g^2-9g^4-3g^2g'^2 \right)(\phi^\dagger \phi)^3 \hs  \\ &+18m^2g^2(\phi^\dagger \phi)^2 -9g^2(\phi^\dagger\phi)\bar{\psi}_L\sqrt{2}H\mathcal{Y}\psi_R  \hs \\ &+g\bar{\psi}_L \sigma^{\mu\nu} W^a_{\mu\nu} \sigma^a \sqrt{2}H\mathcal{Y}\psi_R + \text{h.c.} \bigg) \; . \end{aligned}
\end{align}
Finally, we find the RGE contributions
\begin{align} \begin{aligned}
\beta_{\phi W} &\supseteq \left(6\lambda -\frac{53}{6}g^2 - \frac{3}{2}g'^2 + 2\gamma_\phi \right) C_{\phi W} \; , \qquad &\beta_{\psi\phi} &\supseteq  9g^2 \mathcal{Y} C_{\phi W} \; ,\hs \\ 
\beta_{\phi W B} &\supseteq 2gg' C_{\phi W}\; , &\beta_{\psi W} &\supseteq -g\mathcal{Y} C_{\phi W} \; , \hs \\
\beta_\phi &\supseteq \left(18\lambda g^2-9g^4 - 3g^2 g'^2 \right) C_{\phi W}  \; , &\beta_{g} &\supseteq -8g\frac{m^2}{\Lambda^2} C_{\phi W} \; ,\hs \\
\beta_\lambda &\supseteq 36g^2 \frac{m^2}{\Lambda^2} C_{\phi W} \; . 
\end{aligned}
\end{align}

\subsubsection*{Divergences and RGEs from $\phi B$}

The last operator of the class $\phi X$ is
\begin{align}
Q_{\phi B} = (\phi^\dagger \phi)B_{\mu\nu} B^{\mu\nu} \; .
\end{align}
The fluctuations in (\ref{phiG_fluc}), converted to $U(1)$, lead to
\begin{align}
\begin{aligned}
\text{tr }cM =& - \left(6\lambda +\frac{3}{2}g^2 + \frac{5}{2}g'^2\right) (\phi^\dagger \phi)B_{\mu\nu} B^{\mu\nu}+ 4m^2 B_{\mu\nu} B^{\mu\nu} \\& -2gg' (\phi^\dagger \sigma^a \phi)W^{a}_{\mu\nu} B^{\mu\nu} +3g'^2 (\phi^\dagger\phi) \Box(\phi^\dagger \phi) \; ,  \\
-\frac{1}{4}\text{tr }a^\lambda_{\ \lambda} M^2 =& -\left(6\lambda g'^2  -3g^2g'^2 - 3g'^4\right)(\phi^\dagger \phi)^3-3g'^2 (\phi^\dagger \phi)\Box (\phi^\dagger \phi) \hs \\&+6m^2g'^2(\phi^\dagger \phi)^2  -3g'^2 (\phi^\dagger \phi) \bar{\psi}_L \sqrt{2} H \mathcal{Y} \psi_R + \text{h.c.} \; ,\\
-\frac{1}{12}\text{tr }a^\lambda_{\ \lambda}[D_\mu, [D^\mu, M]  =&\ 2g'^2 (\phi^\dagger \phi)\Box (\phi^\dagger \phi)  \; , \\
\frac{1}{3}\text{tr }a^{\mu\nu}[D_\mu,[D_\nu,M]] =& -2g'^2 (\phi^\dagger \phi)\Box (\phi^\dagger \phi) \; , \\
i\bar{\Gamma}\slashed{b}\Gamma =&\ 2g'\bar{\psi}_L \sigma^{\mu\nu} B_{\mu\nu} \sqrt{2}H\mathcal{Y}(Y_L+Y_R)\psi_R +\text{h.c.} \; , \hs \\ -\frac{1}{6}\text{tr }\bar{\Gamma} i\overset{\text{\footnotesize{$\leftrightarrow$}}}{\mathcal{D}}_{a} \Gamma  =& -24g'^2(\phi^\dagger \phi) \bar{\psi}_L \sqrt{2}H\mathcal{Y} Y_L Y_R \psi_R + \text{h.c.} 
\end{aligned}
\end{align}
Summing all the terms, we obtain
\begin{align}\begin{aligned}
32\pi^2\epsilon \mathcal{L}^\text{div}_{\phi B} = \frac{C_{\phi B}}{\Lambda^2} \bigg( &-\left(6\lambda +\frac{3}{2}g^2 + \frac{5}{2}g'^2 \right)(\phi^\dagger \phi)B_{\mu\nu} B^{\mu\nu} +4m^2B_{\mu\nu} B^{\mu\nu} \\ &-2gg'(\phi^\dagger \sigma^a \phi)W^a_{\mu\nu} B^{\mu\nu} - \left(6\lambda g'^2 -3g^2g'^2 - 3g'^4 \right)(\phi^\dagger \phi)^3 \hs\\ &+6m^2g'^2(\phi^\dagger \phi)^2 -12g'^2(\phi^\dagger \phi) \bar{\psi}_L \sqrt{2}H\mathcal{Y}(Y_L^2+Y_R^2)\psi_R \hs \\ &+2g'\bar{\psi}_L \sigma^{\mu\nu} B_{\mu\nu} \sqrt{2}H\mathcal{Y}(Y_L + Y_R)\psi_R + \text{h.c.} \hs \bigg) \; .
\end{aligned}
\end{align}
The resulting contributions to the RGEs are
\begin{align} \begin{aligned}
\beta_{\phi B} &\supseteq \left(6\lambda -\frac{9}{2}g^2 + \frac{85}{6}g'^2 +2\gamma_\phi \right) C_{\phi B} \; , \qquad &\beta_{\psi\phi} &\supseteq 12g'^2(Y_L^2+Y_R^2)\mathcal{Y}C_{\phi B} \; ,\hs\\
\beta_{\phi W B} &\supseteq 2gg' C_{\phi B} \; , \hs &\beta_{\psi B} &\supseteq -2g'(Y_L + Y_R) \mathcal{Y}  C_{\phi B} \; , \hs \\
\beta_{\phi} &\supseteq \left(6\lambda g'^2 -3g^2g'^2 -3g'^4 \right) C_{\phi B} \; , &\beta_{g'} &\supseteq -8g' \frac{m^2}{\Lambda^2} C_{\phi B} \; ,\hs \\
\beta_{\lambda} &\supseteq 12g'^2\frac{m^2}{\Lambda^2} C_{\phi B} \; .
\end{aligned}
\end{align}

\subsubsection{$\phi\wt{X}$-operators}

The operators of the class $\phi\wt{X}$ mix into CP-violating operators 
that we have not considered yet. We therefore introduce a Lagrangian
\begin{align}
\mathcal{L}_\theta = \frac{\theta_s g_s^2}{32\pi^2} \wt{G}_{\mu\nu}^A G^{A\mu\nu} + 
\frac{\theta g^2}{32\pi^2} \wt{W}_{\mu\nu}^a W^{a\mu\nu} + 
\frac{\theta' g'^2}{32\pi^2} \wt{B}_{\mu\nu} B^{\mu\nu} \label{CP_violating}
\end{align}

Although these operators correspond to total derivatives and play no role in 
perturbation theory, they are related to non-perturbative effects. 
For the beta functions, we will use (\ref{CP_violating}) as reference.

The prototype for operator class $\phi \wt{X}$ is 
\begin{align}
Q_{\phi\wt{G}} = \phi^\dagger \phi~ \wt{G}^A_{\mu\nu} G^{A\mu\nu} \; .
\end{align}
In this case, the fluctuations are given by the four non-trivial entries
\begin{align}
\begin{aligned}
c_{ij} &= -\wt{G}^A_{\mu\nu} G^{A\mu\nu} \delta_{ij} \; , \qquad  
&b^\lambda_{(A\alpha)(B\beta)} &= \epsilon^{\lambda\alpha\beta\mu}\partial_\mu(\vp^2) 
\delta^{AB} \hs \; , \\
c_{(A\alpha)i}&=  2i (D^\mu \vp)_i \wt{G}^A_{\mu\alpha} \; ,  &b^\lambda_{(A\alpha)i} 
&= 2i \wt{G}^{A}_{\lambda\alpha} \vp_i  \hs \; . \label{phiGtilde_fluc}
\end{aligned}
\end{align}

\subsubsection*{Divergences and RGE for $\phi \wt{G}$}
From the fluctuations (\ref{phiGtilde_fluc}) we obtain 
\begin{align}\begin{aligned}
\text{tr } cM &=  -  \left( 6\lambda + \frac{3}{2}g^2 + 
\frac{1}{2} g'^2 \right) (\phi^\dagger \phi)\wt{G}^A_{\mu\nu} G^{A\mu\nu} +
4m^2 \wt{G}^A_{\mu\nu} G^{A\mu\nu} \; ,  \\
\text{tr } i\bar{\Gamma}\slashed{b}\Gamma &= -32ig_s^2 (\phi^\dagger \phi) 
\bar{q}_L \sqrt{2}H\mathcal{Y} q_R +
4ig_s \bar{q}_L \sigma^{\mu\nu} G_{\mu\nu}^A T^A \sqrt{2} H \mathcal{Y} q_R + 
\text{h.c.} \hs \end{aligned}
\end{align}
The divergent Lagrangian is then
\begin{align}\begin{aligned}
32\pi^2\epsilon \mathcal{L}_{\phi\wt{G}}^\text{div} = 
\frac{C_{\phi\wt{G}}}{\Lambda^2} \bigg(&-  \left( 6\lambda + \frac{3}{2}g^2 + 
\frac{1}{2} g'^2 \right) (\phi^\dagger \phi)\wt{G}^A_{\mu\nu} G^{A\mu\nu} +
4m^2 \wt{G}^A_{\mu\nu} G^{A\mu\nu} \\ &-32ig_s^2 (\phi^\dagger \phi) \bar{q}_L 
\sqrt{2}H\mathcal{Y} q_R +4ig_s \bar{q}_L \sigma^{\mu\nu} G_{\mu\nu}^A T^A 
\sqrt{2} H \mathcal{Y} q_R + \text{h.c.} \bigg) \hs \; . 
\end{aligned}
\end{align}

The first term in $\text{tr}\,cM$ 
gives a self-renormalization of  $Q_{\phi \widetilde G}$, while the second term 
renormalizes the QCD $\theta$ term $\mathcal{L}_{\theta_s} =   
\theta_s    g_s^2/(32 \pi^2)  \wt{G}^A_{\mu\nu} G^{A\mu\nu}$. The two 
contributions in $\text{tr } i\bar{\Gamma}\slashed{b}\Gamma$ already 
correspond to operators of the Warsaw basis.   The term 
$(\phi^\dagger \phi) \bar{q}_L \sqrt{2}H\mathcal{Y} q_R$ corresponds to the 
operators $Q_{u \phi} = (\phi^{\dag} \phi)   (\bar  q_L  u_R \wt{\phi}) $ and 
$Q_{d \phi} = (\phi^{\dag} \phi)   (\bar q_L  d_R \phi)$.  The last piece 
$\bar{q}_L \sigma^{\mu\nu} G_{\mu\nu}^A T^A \sqrt{2} H \mathcal{Y} q_R$ 
corresponds to the operators 
$Q_{uG} = ( \bar q_L \sigma^{\mu \nu}   T^{A}   u_R  )  \wt{\phi}  G_{\mu \nu}^{A}$ 
and $Q_{dG}= ( \bar q_L \sigma^{\mu \nu}   T^{A}   d_R  )  \phi  G_{\mu \nu}^{A}  $.

We deduce the renormalization group contributions 
\begin{align}
\begin{aligned}
\beta_{\phi\wt{G}} &\supseteq \left(6\lambda -14g_s^2 - \frac{9}{2}g^2 -
\frac{3}{2}g'^2 + 2\gamma_\phi \right) C_{\phi\wt{G}} \; , \qquad 
&\beta_{q\phi} &\supseteq 32ig_s^2 \mathcal{Y}_q C_{\phi\wt{G}}\; , \hs \\
\beta_{\theta_s} &\supseteq -\frac{128\pi^2}{g_s^2} \frac{m^2}{\Lambda^2} 
C_{\phi\wt{G}} &\beta_{q G} &\supseteq -4ig_s \mathcal{Y}_q C_{\phi\wt{G}}\hs \; .
\end{aligned}
\end{align}

\subsubsection*{Divergences and RGEs from $\phi \wt{W}$}

The next operator we consider is
\begin{align}
Q_{\phi\wt{W}} = (\phi^\dagger \phi) \wt{W}^a_{\mu\nu} W^{a\mu\nu} \; .
\end{align}
The $SU(2)$-analogy of (\ref{phiGtilde_fluc}) leads to
\begin{align} \begin{aligned}
\text{tr } cM =& - \left( 6\lambda + \frac{7}{2}g^2 + \frac{1}{2} g'^2 \right) 
(\phi^\dagger \phi)\wt{W}^a_{\mu\nu} W^{a\mu\nu} +4m^2 \wt{W}^a_{\mu\nu} W^{a\mu\nu}  \\
&-2gg'(\phi^\dagger \sigma^a \phi)\wt{W}^a_{\mu\nu} B^{\mu\nu} \; , \\
\text{tr } i\bar{\Gamma}\slashed{b}\Gamma =& -9ig^2 (\phi^\dagger \phi) 
\bar{\psi}_L \sqrt{2} H \mathcal{Y} \psi_R + ig\bar{\psi}_L \sigma^{\mu\nu} 
W^a_{\mu\nu} \sigma^a \sqrt{2} H\mathcal{Y} \psi_R + \text{h.c.} \hs
\end{aligned}
\end{align}
Adding these results, we obtain
\begin{align}\begin{aligned}
32\pi^2\epsilon \mathcal{L}_{\phi\wt{W}}^\text{div} = 
\frac{C_{\phi\wt{W}}}{\Lambda^2} \bigg(&-  \left( 6\lambda + \frac{7}{2}g^2 + 
\frac{1}{2} g'^2 \right) (\phi^\dagger \phi)\wt{W}^a_{\mu\nu} W^{a\mu\nu} +
4m^2 \wt{W}^a_{\mu\nu} W^{a\mu\nu} \\&-2gg'(\phi^\dagger \sigma^a \phi)
\wt{W}^a_{\mu\nu} B^{\mu\nu} \hs \\&-9ig^2 (\phi^\dagger \phi) \bar{\psi}_L 
\sqrt{2}H\mathcal{Y} \psi_R +ig \bar{\psi}_L \sigma^{\mu\nu} W_{\mu\nu}^a \sigma^a 
\sqrt{2} H \mathcal{Y} \psi_R + \text{h.c.} \bigg) \hs \; . 
\end{aligned}
\end{align}
The contributions to the beta functions are
\begin{align} \begin{aligned}
\beta_{\phi\wt{W}} &\supseteq \left(6\lambda -\frac{53}{6}g^2 - \frac{3}{2}g'^2 + 
2\gamma_\phi \right) C_{\phi\wt{W}} \; , \qquad 
&\beta_{\psi \phi} &\supseteq 9ig^2 \mathcal{Y} C_{\phi\wt{W}} \hs \; , \\
\beta_{\phi \wt{W} B} &\supseteq 2gg' C_{\phi\wt{W}} \hs \; , &\beta_{\psi W} 
&\supseteq -ig\mathcal{Y} C_{\phi\wt{W}} \; , \hs \\
\beta_\theta &\supseteq -\frac{128\pi^2}{g^2}\frac{m^2}{\Lambda^2} C_{\phi\wt{W}} 
\; .
\end{aligned}
\end{align}

\subsubsection*{Divergences and RGEs from $\phi\wt{B}$}

The last operator of this class is
\begin{align}
Q_{\phi\wt{B}} = (\phi^\dagger \phi) \wt{B}_{\mu\nu} B^{\mu\nu} \; .
\end{align}
We translate the fluctuations (\ref{phiGtilde_fluc}) to the case of $U(1)$ 
and find
\begin{align} \begin{aligned}
\text{tr }cM =& - \left(6\lambda +\frac{3}{2}g^2 +\frac{5}{2}g'^2 \right) 
(\phi^\dagger \phi)\wt{B}_{\mu\nu}B^{\mu\nu} + 4m^2 \wt{B}_{\mu\nu} B^{\mu\nu} \\
& - 2gg' (\phi^\dagger \sigma^a \phi)\wt{W}^a_{\mu\nu} B^{\mu\nu}\; , \\
\text{tr }i\bar{\Gamma}\slashed{b}\Gamma =& -12ig'^2 (\phi^\dagger \phi) 
\bar{\psi}_L \sqrt{2} H \mathcal{Y} (Y^2_L+Y^2_R)\psi_R \hs \\ 
&+ 2ig' \bar{\psi}_L \sigma^{\mu\nu} B_{\mu\nu} \sqrt{2}H\mathcal{Y}(Y_L+Y_R) 
\psi_R + \text{h.c.}
\end{aligned}
\end{align}
These results add up to the divergent Lagrangian
\begin{align}\begin{aligned}
32\pi^2\epsilon \mathcal{L}_{\phi\wt{B}}^\text{div} = 
\frac{C_{\phi\wt{B}}}{\Lambda^2} \bigg(&-  \left( 6\lambda + \frac{3}{2}g^2 + 
\frac{5}{2} g'^2 \right) (\phi^\dagger \phi)\wt{B}_{\mu\nu} B^{\mu\nu} +
4m^2 \wt{B}_{\mu\nu} B^{\mu\nu} \\&-2gg'(\phi^\dagger \sigma^a \phi)\wt{W}^a_{\mu\nu} 
B^{\mu\nu} \hs -12ig'^2 (\phi^\dagger \phi) \bar{\psi}_L \sqrt{2}H
\mathcal{Y}(Y_L^2+Y_R^2) \psi_R \\&+2ig' \bar{\psi}_L \sigma^{\mu\nu} B_{\mu\nu} 
\sqrt{2} H \mathcal{Y}(Y_L+Y_R) \psi_R + \text{h.c.} \bigg) \hs \; . 
\end{aligned}
\end{align}
We infer the following contributions to the RGEs 
\begin{align}
\begin{aligned}
\beta_{\phi\wt{B}} &\supseteq \left(6\lambda -\frac{9}{2}g^2 +\frac{85}{6}g'^2 + 
2\gamma_\phi \right) C_{\phi\wt{B}} \; , \quad & \beta_{\psi\phi} 
&\supseteq 12ig'^2 (Y_L^2+Y_R^2) \mathcal{Y}  C_{\phi \wt{B}}\; , \\
\beta_{\phi\wt{W}B} &\supseteq 2gg' C_{\phi\wt{B}}\; , 
&\beta_{\psi B} &\supseteq -2ig'(Y_L+Y_R) \mathcal{Y}  C_{\phi \wt{B}} \; ,\hs \\
\beta_{\theta'} 
&\supseteq -\frac{128\pi^2}{g'^2}\frac{m^2}{\Lambda^2}C_{\phi\wt{B}} \; .
\end{aligned}
\end{align}

\subsubsection{$\phi W B$-operator}

This operator is given by
\begin{align}
Q_{\phi W B} = (\phi^\dagger \sigma^a \phi)W_{\mu\nu}^a B^{\mu\nu} =  
 2(\vp t^a_{L}t^3_{R}\vp) W_{\mu\nu}^a B^{\mu\nu}
\end{align}
The quadratic fluctuations are described by the matrices
\begin{align}\begin{aligned}
c_{(a\alpha)(b\beta)} =& -4g\epsilon^{abc} (\vp t_L^c t_R^3\vp) B_{\alpha\beta} \;, 
\qquad  
\hs &a^{\mu\nu}_{(a\alpha)\beta} =&-2 (\vp t_L^a t_R^3 \vp) S^{\alpha\beta\mu\nu} \; , \\
c_{(a\alpha)\beta} =&\ 2(D_\mu\vp t_L^a t_R^3 D_\nu \vp) S^{\alpha\beta\mu\nu} 
&b^\lambda_{(a\alpha)\beta} =&\ 2g^{\alpha\lambda} (\vp t_L^a t_R^3 D_\beta \vp) \hs \\
&+2(\vp t_L^at_R^3 D_\mu D_\nu \vp) S^{\alpha\beta\mu\nu} &&-2g^{\beta\lambda} 
(\vp t_L^a t_R^3 D_\alpha \vp) \; ,\\
&+ g\epsilon^{abc} W_{\alpha\beta}^b (\vp t_L^c t_R^3\vp) \; , \\
c_{(a\alpha)i} =&\ 4i(t_L^a t_R^3 \vp)_i \partial^\mu B_{\mu\alpha} + 
4i (t_L t_R^3 D^\mu \vp) B_{\mu\alpha} \; , 
&b^\lambda_{(a\alpha)i} =&\ 4i(t_L^a t_R^3 \vp)_i B^{\lambda\alpha} \hs \; , \\
c_{\alpha i} =&\ 4i (t_L^a t_R^3 \vp)_i D^\mu W^a_{\mu\alpha} + 
4i (t_L^a t_R^3 D^\mu \vp) W^a_{\mu\alpha} \; , &b^\lambda_{\alpha i} 
&= 4i(t_L^a t_R^3\vp)W^{a\lambda\alpha} \; ,\hs \\
c_{ij} =& -4(t_L^a t_R^3)_{ij} W^a_{\mu\nu} B^{\mu\nu} \; . \hs 
\end{aligned}
\end{align}
The symmetric tensor $S^{\alpha\beta\mu\nu}$ has been defined below 
eq.~(\ref{phiG_fluc}). We obtain the terms
\begin{align}\begin{aligned}
\text{tr }cM =& -\left(2\lambda + \frac{5}{2}g^2 + 4g^2C_2^\text{ad}+
\frac{3}{2}g'^2 \right) (\phi^\dagger \sigma^a \phi)W^a_{\mu\nu} B^{\mu\nu} \\
&-gg'(\phi^\dagger \phi)W^a_{\mu\nu} W^{a\mu\nu} 
-3gg'(\phi^\dagger \phi)B_{\mu\nu} B^{\mu\nu} -6\lambda gg' (\phi^\dagger \phi)^3 \\ 
& + 24gg' (D_\mu \vp t_L^a t_R^3 D^\mu \vp)(\vp t_L^a t_R^3 \vp) 
+ 6m^2 gg'(\phi^\dagger \phi)^2  \\ 
&-3gg' (\phi^\dagger \phi) \bar{\psi}_L \sqrt{2}H\mathcal{Y} \psi_R + 
\text{h.c.} \; ,  \\
-\frac{1}{4}\text{tr }a^\lambda_{\ \lambda} M^2 =& \left(3g^3g'+3gg'^3 \right) 
(\phi^\dagger \phi)^3 -24gg'(D_\mu \vp t_L^a t_R^3 D^\mu \vp)(\vp t_L^a t_R^3 \vp) 
\; , \\ -\frac{1}{12} \text{tr }a^\lambda_{\ \lambda}[D_\mu,[D^\mu,M]] 
=&\ 16gg' (\vp t_L^a t_R^3 D_\mu D^\mu \vp + D_\mu \vp t_L^a t_R^3 D^\mu \vp)
(\vp t_L^a t_R^3 \vp) \; , \\ \frac{1}{3}\text{tr }a^{\mu\nu}[D_\mu,[D_\nu,M]] 
=& -16gg' (\vp t_L^a t_R^3 D_\mu D^\mu \vp + D_\mu \vp t_L^a t_R^3 D^\mu \vp)
(\vp t_L^a t_R^3 \vp) \; , \\ \text{tr }i\bar{\Gamma} \slashed{b} \Gamma 
=& -2g' \bar{\psi}_L \sigma^{\mu\nu} W^a_{\mu\nu} \sigma^a \sqrt{2}H \mathcal{Y} 
(Y_L + Y_R)\tau^3\psi_R \hs \\ 
&-3g\bar{\psi}_L \sigma^{\mu\nu} B_{\mu\nu} \sqrt{2} H \mathcal{Y}\tau^3 
\psi_R + \text{h.c.} \; , \\ -\frac{1}{6}\text{tr }\bar{\Gamma} 
i\overset{\text{\footnotesize{$\leftrightarrow$}}}{\mathcal{D}}_{a} \Gamma 
=&\ 12gg'(\phi^\dagger \phi) \bar{\psi}_L \sqrt{2} H \mathcal{Y} Y_R \tau^3 
\psi_R + \text{h.c.} 
\end{aligned}
\end{align}
Adding these terms leads to
\begin{align}\begin{aligned}
32\pi^2\epsilon \mathcal{L}^\text{div}_{\phi W B} = \frac{C_{\phi W B}}{\Lambda^2} 
\bigg( &-\left(2\lambda +\frac{21}{2}g^2 +\frac{3}{2}g'^2 \right) 
(\phi^\dagger\sigma^a\phi)W^a_{\mu\nu} B^{\mu\nu} + 6m^2gg'(\phi^\dagger\phi)^2 \\
&-\left(6\lambda gg' -3g^3g' -3gg'^3 \right)(\phi^\dagger \phi)^3 -
gg' (\phi^\dagger \phi)W^a_{\mu\nu} W^{a\mu\nu} \hs \\&-3gg'(\phi^\dagger \phi) 
B_{\mu\nu} B^{\mu\nu} +12gg'(\phi^\dagger \phi) \bar{\psi}_L \sqrt{2}H\mathcal{Y} 
Y_L \tau^3\psi_R \hs \\ &-2g'\bar{\psi}_L \sigma^{\mu\nu} W^a_{\mu\nu} \sigma^a 
\sqrt{2}H\mathcal{Y}(Y_L+Y_R)\tau^3\psi_R \hs \\& -3g\bar{\psi}_L\sigma^{\mu\nu} 
B_{\mu\nu} \sqrt{2}H\mathcal{Y}\tau^3\psi_R +\text{h.c.} \hs \bigg) \; .
\end{aligned}
\end{align}
We find eight contributions to the beta functions:
\begin{align} \begin{aligned}
\beta_{\phi W B} &\supseteq \left(2\lambda +\frac{4}{3}g^2+\frac{19}{3}g'^2+
2\gamma_\phi \right)C_{\phi W B} \; , \quad &\beta_{\psi\phi}&\supseteq 
-12gg'Y_L\tau^3\mathcal{Y}C_{\phi W B} \; , \\ \beta_{\phi W} 
&\supseteq gg'C_{\phi W B} \; , &\beta_{\psi W}&\supseteq 2g'(Y_L+Y_R)\tau^3
\mathcal{Y} C_{\phi W B} \; , \hs \\ \beta_{\phi B} &\supseteq 3gg' C_{\phi W B} \; ,
&\beta_{\psi B} &\supseteq 3g\tau^3 \mathcal{Y} C_{\phi W B} \; , \hs \\ 
\beta_\phi &\supseteq \left(6\lambda gg' -3g^3g' -3gg'^3 \right)C_{\phi W B} \; , 
&\beta_\lambda &\supseteq 12gg' \frac{m^2}{\Lambda^2} C_{\phi W B} \; .
\end{aligned}
\end{align}

\subsubsection{$\phi \wt{W} B$-operator}

We are left with the last operator of class $X^2\phi^2$, 
\begin{align}
Q_{\phi\wt{W}B}  = (\phi^\dagger \sigma^a \phi)\wt{W}_{\mu\nu}^a B^{\mu\nu} 
= 2(\vp t^a_{L}t^3_{R}\vp)\widetilde{W}_{\mu\nu}^a B^{\mu\nu} \; .
\end{align}
The non-trivial elements of the fluctuation matrices are
\begin{align}
\begin{aligned}
c_{(a\alpha)(b\beta)} &= -4g\epsilon^{abc}(\vp t_L^c t_R^3 \vp)\wt{B}_{\alpha\beta} \; , 
\qquad &b^\lambda_{(a\alpha)\beta}&=-4\epsilon^{\alpha\beta\lambda\mu} 
(\vp t_L^a t_R^3 D_\mu \vp) \; , \hs \\
c_{(a\alpha)\beta} &= 4ig (\vp t_L^a t_R^3 t_L^b \vp) \wt{W}^b_{\alpha\beta}\; , 
&b^\lambda_{(a\alpha)i}&= 4i (t_L^a t_R^3 \vp)_i \wt{B}^{\lambda\alpha} \; , \hs \\
c_{(a\alpha)i} &= 4i (t_L^a t_R^3 D^\mu \vp)_i \wt{B}_{\mu\alpha} \; , \qquad 
&b^\lambda_{\alpha i}&= 4i(t_L^a t_R^3 \vp)_i \wt{W}^{a\lambda\alpha} \; , \hs \\
c_{\alpha i} &=4i (t_L^a t_R^3 D^\mu \vp)_i \wt{W}^a_{\mu\alpha} \; , \hs  \\
c_{ij} &= -4 (t_L^a t_R^3)_{ij} \wt{W}^a_{\mu\nu} B^{\mu\nu} \; . \hs
\end{aligned}
\end{align}
From these, we calculate the divergences
\begin{align}
\begin{aligned}
\text{tr }cM &= -\left(2\lambda +\frac{5}{2}g^2 + 4g^2C_2^\text{ad} + 
\frac{3}{2}g'^2 \right) (\phi^\dagger \sigma^a \phi)\wt{W}^a_{\mu\nu}B^{\mu\nu} \\
&\hphantom{=} -gg' (\phi^\dagger \phi)\wt{W}^a_{\mu\nu} W^{\mu\nu} 
-3gg' (\phi^\dagger \phi)\wt{B}_{\mu\nu} B^{\mu\nu}  \; , \\
\text{tr }i\bar{\Gamma}\slashed{b}\Gamma &=12igg'(\phi^\dagger \phi) 
\bar{\psi}_L \sqrt{2}H\mathcal{Y} Y_L \tau^3 \psi_R -3ig\bar{\psi}_L
\sigma^{\mu\nu}B_{\mu\nu} \sqrt{2}H\mathcal{Y}\tau^3\psi_R  \hs \\ 
&\hphantom{=}-2g'\bar{\psi}_L\sigma^{\mu\nu} W^a_{\mu\nu} \sigma^a 
\sqrt{2}H\mathcal{Y} (Y_L+Y_R) \tau^3 \psi_R+ \text{h.c.}
\end{aligned}
\end{align}
Adding these results, the divergent Lagrangian becomes
\begin{align} \begin{aligned}
32\pi^2\epsilon \mathcal{L}^\text{div}_{\phi \wt{W} B} = 
\frac{C_{\phi \wt{W} B}}{\Lambda^2} \bigg(&-\left(2\lambda +\frac{21}{2}g^2 + 
\frac{3}{2}g'^2 \right)(\phi^\dagger \sigma^a \phi)\wt{W}^a_{\mu\nu} B^{\mu\nu} \\
&-gg'(\phi^\dagger \phi)\wt{W}^a_{\mu\nu}W^{a\mu\nu} +12igg'(\phi^\dagger \phi) 
\bar{\psi}_L \sqrt{2}H\mathcal{Y} Y_L \tau^3 \psi_R \hs \\
&-3gg'(\phi^\dagger \phi)\wt{B}_{\mu\nu} B^{\mu\nu} 
-3ig\bar{\psi}_L\sigma^{\mu\nu}B_{\mu\nu} \sqrt{2}H\mathcal{Y}\tau^3\psi_R \hs \\ 
&-2g'\bar{\psi}_L\sigma^{\mu\nu} W^a_{\mu\nu} \sigma^a \sqrt{2}H\mathcal{Y} 
(Y_L+Y_R) \tau^3 \psi_R +\text{h.c.} \bigg) \hs \; .
\end{aligned} 
\end{align}
The contributions to the RGEs are then
\begin{align}\begin{aligned}
\beta_{\phi \wt{W} B} &\supseteq \left(2\lambda +\frac{4}{3}g^2 + 
\frac{19}{3}g'^2+2\gamma_\phi \right) C_{\phi \wt{W} B} \; , \quad 
& \beta_{\psi\phi}&\supseteq-12igg'Y_L\tau^3 \mathcal{Y}C_{\phi \wt{W} B}  \;, \\
\beta_{\phi \wt{W}} &\supseteq gg' C_{\phi \wt{W} B} \; , &\beta_{\psi W}
&\supseteq 2ig'(Y_L+Y_R)\tau^3\mathcal{Y}C_{\phi \wt{W} B} \; , \hs \\
\beta_{\phi \wt{B}} &\supseteq 3gg' C_{\phi \wt{W} B} \; , &\beta_{\psi B}
&\supseteq 3ig\tau^3\mathcal{Y} C_{\phi \wt{W} B} \; . \hs 
\end{aligned}
\end{align}

\subsection{Renormalization of the operator class $\phi^6$ 
and $\phi^4D^2$}

This class collects the operators that consist only of scalar fields and 
derivatives, namely $Q_\phi$, $Q_{\phi\Box}$ and $Q_{\phi D}$. 
Using the relations in appendix \ref{sec:app2}, these operators 
can be expressed in terms of the four real scalar fields $\varphi_i$. 
The basic building blocks in this representation are
$(\vp^2)^3$,  $(\vp D_\mu\vp)^2$ and $(\vp t_R^3 D_\mu \vp)^2$.

\subsubsection{$\phi^6$-operator}

We begin with the simplest case  
\begin{align}
Q_{\phi}=(\phi^\dagger\phi)^3 = \frac{1}{8} (\varphi_i \varphi_i)^3 \; .
\end{align}
There is only one non-trivial matrix entry:
\begin{align}
c_{ij} = - \frac{3}{4}(\vp^2)^2\delta_{ij} - 3(\vp^2)\vp_i\vp_j \; .
\end{align}
From the master formula we find
\begin{align}
\text{tr }cM = -\left(54\lambda+\frac{9}{2}g^2 + \frac{3}{2}g'^2 \right)
(\phi^\dagger\phi)^3 + 24m^2(\phi^\dagger\phi)^2 \; .
\end{align}
The divergent Lagrangian is then given by
\begin{align}
32\pi^2\epsilon\mathcal{L}_{\phi}^\text{div} = \frac{C_{\phi}}{\Lambda^2} 
\left( -\left(54\lambda+\frac{9}{2}g^2 + \frac{3}{2}g'^2 \right)
(\phi^\dagger\phi)^3 + 24m^2(\phi^\dagger\phi)^2 \right) \; .
\end{align}
We obtain the two beta-function contributions
\begin{align}\begin{aligned}
\beta_\phi \supseteq \left(54\lambda  -\frac{27}{2}g^2 - \frac{9}{2}g'^2 + 
6\gamma_\phi \right)C_{\phi} \; ,\qquad 
\beta_\lambda \supseteq 48\frac{m^2}{\Lambda^2} C_{\phi} \; .
\end{aligned}
\end{align}

\subsubsection{$\phi\Box$-operator}
For the next operator, we have
\begin{align}
Q_{\phi\Box} = (\phi^\dagger\phi)\Box(\phi^\dagger\phi) = -(\vp D_\mu \vp)^2 \; ,
\end{align}
dropping a total derivative. The fluctuation matrices are
\begin{align}
\begin{aligned}
c_{ij} &= 2(D_\mu \vp)_i (D^\mu \vp)_j - \Box(\vp^2) \delta_{ij} \; , \hs\\
b^\mu_{ij} &= (D^\mu \vp)_i \vp_j - \vp_i (D^\mu \vp)_j  \; , \hs \\
a^{\mu\nu}_{ij} &= -2 \vp_i \vp_j g^{\mu\nu} \equiv a_{ij}g^{\mu\nu}\; . \hs
\end{aligned}
\end{align}
From these, we find the terms of the master formula to be
\begin{align}
\begin{aligned}
\text{tr }cM =&-\left( 12\lambda + \frac{3}{2}g^2 + \frac{1}{2}g'^2\right)(\phi^\dagger \phi)\Box(\phi^\dagger \phi) +2g'^2(\phi^\dagger D_\mu \phi)^*(\phi^\dagger D^\mu \phi) \\ &+\left(4\lambda^2+2\lambda g^2 \right)(\phi^\dagger \phi)^3  +4m^2(\phi^\dagger \phi) -\left(8\lambda+2g^2 \right)m^2 (\phi^\dagger \phi)^2 \\ &-2m^2\bar{\psi}_L\sqrt{2}H\mathcal{Y}\psi_R + (2\lambda+g^2)(\phi^\dagger \phi) \bar{\psi}_L \sqrt{2}H\mathcal{Y} \psi_R + \text{h.c.}  \; , \\ -\frac{1}{3}\text{tr }[D^\mu, N_{\mu\nu}]b^\nu =& -\left(\frac{1}{2}g^2 + \frac{1}{6}g'^2 \right)(\phi^\dagger \phi)\Box(\phi^\dagger \phi) -\frac{2}{3}g'^2(\phi^\dagger D_\mu \phi)^*(\phi^\dagger D^\mu \phi) \\ &-\frac{2}{3}\lambda g^2 (\phi^\dagger \phi)^3 + \frac{2}{3}g^2m^2(\phi^\dagger \phi)^2 - \frac{1}{3}g^2 (\phi^\dagger \phi)\bar{\psi}_L \sqrt{2}H\mathcal{Y}\psi_R \\ &-\frac{1}{3}g'^2 (\phi^\dagger i \overset{\text{\footnotesize$\leftrightarrow$}}{D_\mu} \phi) \bar{\psi}\gamma^\mu(Y_LP_L+Y_RP_R)\psi \\&-\frac{1}{6}g^2 (\phi^\dagger i\overset{\text{\footnotesize$\leftrightarrow$}}{D^a_\mu} \phi) \bar{\psi}_L \gamma^\mu \sigma^a \psi_L + \text{h.c.} \; , \\ -\text{tr }aM^2 =& -\left(6g^2+2g'^2 \right)(\phi^\dagger \phi)\Box(\phi^\dagger \phi) -8g'^2(\phi^\dagger D_\mu \phi)^*(\phi^\dagger D^\mu \phi) \hs \\ &+\left(36\lambda^2 -8\lambda g^2\right) (\phi^\dagger \phi)^3 -\left(24\lambda -8g^2 \right)m^2(\phi^\dagger \phi)^2 \\ &+4m^4(\phi^\dagger \phi) -4g^2 (\phi^\dagger \phi) \bar{\psi}_L \sqrt{2}H\mathcal{Y} \psi_R + \text{h.c.} \; , \\ \text{tr }i\bar{\Gamma}\slashed{b}\Gamma =&\ \frac{1}{2}(\phi^\dagger i \overset{\text{\footnotesize{$\leftrightarrow$}}}{D^a_\mu}\phi)\bar{\psi}_L\gamma^\mu \sigma^a \langle\mathcal{Y}\mathcal{Y}^\dagger \rangle_I \psi_L +2(\phi^\dagger i\overset{\text{\footnotesize{$\leftrightarrow$}}}{D_\mu}\phi) \bar{\psi}_R \gamma^\mu \mathcal{Y}^\dagger \mathcal{Y} \tau^3 \psi_R \hs  \\ &+(\phi^\dagger i\overset{\text{\footnotesize{$\leftrightarrow$}}}{D_\mu}\phi)\bar{\psi}_L\gamma^\mu \left(\langle \mathcal{Y} \mathcal{Y}^\dagger \rangle_I - 2\mathcal{Y} \mathcal{Y}^\dagger \right)\tau^3\psi_L\\ &+4i\bar{\psi}_R \gamma^\mu \mathcal{Y}^\dagger \vp(\tau^1 t_R^1 + \tau^2 t_R^2)D_\mu \vp \mathcal{Y} \psi_R \; ,
\\-\frac{1}{6}\text{tr }\bar{\Gamma} i\overset{\text{\footnotesize{$\leftrightarrow$}}}{\mathcal{D}}_{a} \Gamma  &= 6(\phi^\dagger\phi) \bar{\psi}_L \sqrt{2}H\mathcal{Y}\mathcal{Y}^\dagger\mathcal{Y} \psi_R + \text{h.c.} 
\end{aligned}
\end{align}
The results sum up to the divergent Lagrangian
\begin{align}
\begin{aligned}
32\pi^2\epsilon\mathcal{L}^\text{div}_{\phi\Box} = \frac{C_{\phi\Box}}{\Lambda^2}\bigg( &-\left(12\lambda + 8g^2+\frac{8}{3}g'^2\right)(\phi^\dagger\phi)\Box(\phi^\dagger \phi) -\frac{20}{3}g'^2(\phi^\dagger D_\mu \phi)^*(\phi^\dagger D^\mu \phi) \\ &+\left(40\lambda^2-\frac{20}{3}\lambda g^2 \right)(\phi^\dagger \phi)^3 -\left(32\lambda-\frac{20}{3}g^2\right)m^2(\phi^\dagger\phi)^2 \\&+ 8m^4(\phi^\dagger\phi) +(\phi^\dagger\phi)\bar{\psi}_L\sqrt{2}H\left(2\lambda\mathcal{Y} -\frac{10}{3}g^2\mathcal{Y} + 6\mathcal{Y}\mathcal{Y}^\dagger \mathcal{Y} \right)\psi_R \\&-2m^2\bar{\psi}_L\sqrt{2}H\mathcal{Y}\psi_R  -(\phi^\dagger i\overset{\text{\footnotesize{$\leftrightarrow$}}}{D_\mu}\phi)\bar{\psi}_R \gamma^\mu \left(\frac{1}{3}g'^2Y_R -2\mathcal{Y}^\dagger \mathcal{Y}\tau^3 \right) \psi_R \\ &-(\phi^\dagger i\overset{\text{\footnotesize{$\leftrightarrow$}}}{D_\mu}\phi)\bar{\psi}_L\gamma^\mu \left( \frac{1}{3}g'^2Y_L +\left(2\mathcal{Y}\mathcal{Y}^\dagger -\langle\mathcal{Y}\mathcal{Y}^\dagger \rangle_I\right)\tau^3 \right) \psi_L \\ &-(\phi^\dagger i\overset{\text{\footnotesize{$\leftrightarrow$}}}{D^a_\mu}\phi)\bar{\psi}_L\gamma^\mu \sigma^a\left( \frac{1}{6}g^2-\frac{1}{2}\langle\mathcal{Y}\mathcal{Y}^\dagger \rangle_I \right) \psi_L \\ &+4i\bar{\psi}_R\gamma^\mu \mathcal{Y}^\dagger \vp(\tau^1 t_R^1 + \tau^2 t_R^2)D_\mu \vp\mathcal{Y}\psi_R +\text{h.c.} \bigg) \; . \label{box_div}
\end{aligned}
\end{align}
In total, the $\phi\Box$-operator contributes to the beta-functions with
\begin{align}
\begin{aligned}
\beta_{\phi\Box} &\supseteq \left(12\lambda -4g^2-\frac{4}{3}g'^2+4\gamma_\phi \right)C_{\phi\Box} \; , \quad &\beta_{\psi \phi} &\supseteq\left(-2\lambda \mathcal{Y} +\frac{10}{3}g^2\mathcal{Y} -6\mathcal{Y}\mathcal{Y}^\dagger \mathcal{Y} \right) C_{\phi\Box} \; ,\\
\beta_{\phi D} &\supseteq \frac{20}{3}g'^2 C_{\phi\Box}\; , &\beta_{\phi\psi} &\supseteq \left(\frac{1}{3}g'^2Y_R-2\tau^3\mathcal{Y}^\dagger\mathcal{Y}\right) C_{\phi\Box} \; , \\
\beta_\phi &\supseteq \left(-40\lambda^2+\frac{20}{3}\lambda g^2 \right) C_{\phi\Box} \; , &\beta^{(1)}_{\phi\psi}&\supseteq \left(\frac{1}{3}g'^2Y_L +\left(2\mathcal{Y}\mathcal{Y}^\dagger -\langle\mathcal{Y}\mathcal{Y}^\dagger \rangle_I\right)\tau^3 \right)C_{\phi\Box} \; , \\
\beta_\lambda &\supseteq  \left(-64\lambda+\frac{40}{3}g^2 \right) \frac{m^2}{\Lambda^2} C_{\phi\Box} \; , &\beta^{(3)}_{\phi\psi}&\supseteq \left(\frac{1}{6}g^2-\frac{1}{2}\langle\mathcal{Y}\mathcal{Y}^\dagger \rangle_I \right) C_{\phi\Box} \; ,\\
\beta_{m^2} &\supseteq  -8\frac{m^4}{\Lambda^2}C_{\phi\Box} \; , &\beta_{\phi ud} &\supseteq 2\mathcal{Y}^\dagger_u \mathcal{Y}_d C_{\phi\Box} \; , \\
\beta_\mathcal{Y} &\supseteq -2\frac{m^2}{\Lambda^2}\mathcal{Y}C_{\phi\Box} \; . \label{beta_phibox}
\end{aligned}
\end{align}

\subsubsection{$\phi D$-operator}

The $\phi D$ operator can be decomposed as
\begin{align}\label{boxdr}
Q_{\phi D} = -\frac{1}{4} Q_{\phi\Box} + Q_{\phi R} \; .
\end{align}
where
\begin{align}
Q_{\phi R} = - (\vp t_R^3 D_\mu \vp)^2 \; .
\end{align}
$Q_{\phi\Box}$ has already been treated above. 
For $Q_{\phi R}$ we find the fluctuations
\begin{align} \begin{aligned}
a^{\mu\nu}_{ij} &= -2 (t_R^3\vp)_i (t_R^3\vp)_j g^{\mu\nu} \equiv a_{ij}g^{\mu\nu} \; , \hs \\
b^\lambda_{ij} &= 2 (\vp t_R^3 D^\lambda \vp) t_{Rij}^3 + (t_R^3\vp)_i (t_R^3 D^\lambda\vp)_j - (t_R^3 D^\lambda\vp)_i (t_R^3\vp)_j \hs \; , \\
b^\lambda_{(a\alpha)i} &= g(\vp t_L^a t_R^3 \vp) (t_R^3\vp)_i g^{\lambda \alpha} \; , \hs \\
b^\lambda_{\alpha i} &= \frac{g'}{4} \vp^2 (t_R^3\vp)_i g^{\lambda\alpha} \; , \\
c_{ij} &= 6 (t_R^3 D_\mu \vp)_i (t_R^3 D^\mu \vp)_j + 2(t_R^3\vp)_i (t_R^3 D^2\vp)_j + 2(t_R^3 D^2\vp)_i(t_R^3\vp)_j \; , \hs \\
c_{(a\alpha)(b\beta)} &= 2g^2(\vp t_L^a t_R^3 \vp)(\vp t_L^b t_R^3 \vp) g_{\alpha\beta} \; , \hs \\
c_{\alpha\beta} &= \frac{g'^2}{8} (\vp^2)^2 g_{\alpha\beta} \; , \\
c_{(a\alpha)\beta} &= \frac{gg'}{2} \vp^2 (\vp t_L^a t_R^3 \vp) g_{\alpha\beta} \; , \hs \\
c_{(a\alpha)i} &= -4g (\vp t_R^3 D_\alpha \vp) (t_L^a t_R^3 \vp)_i -3g (\vp t_L^a t_R^3 \vp) (t_R^3 D_\alpha \vp)_i -2g (\vp t_L^a t_R^3 D_\alpha \vp) (t_R^3 \vp)_i \; , \hs \\
c_{\alpha i} &= -g'(\vp t_R^3 D_\alpha \vp) \vp_i - \frac{3}{4} g' \vp^2 (t_R^3 D_\alpha \vp)_i - \frac{g'}{4} \partial_\alpha (\vp^2) (t_R^3\vp)_i \; .  
\end{aligned}
\end{align}
The divergences derived from these matrices are
\begin{align}
\begin{aligned}
\text{tr }cM =& -\left(6\lambda+\frac{29}{2}g^2+4g'^2\right) (\phi^\dagger D_\mu \phi)^*(\phi^\dagger D^\mu \phi) \\ &-\left( 3\lambda+\frac{19}{8}g^2+\frac{13}{8}g'^2\right)(\phi^\dagger \phi)\Box(\phi^\dagger \phi) -m^4(\phi^\dagger \phi) \\ &-\left(\lambda^2 - \frac{1}{2}\lambda g^2 + 2\lambda g'^2 -g^4 -2g^2 g'^2 - g'^4 \right) (\phi^\dagger \phi)^3  \\ &+\left(2\lambda -\frac{1}{2}g^2 +2g'^2 \right)m^2(\phi^\dagger \phi)^2 +\frac{1}{2}m^2\bar{\psi}_L \sqrt{2}H\mathcal{Y}\psi_R  \\ & -\left(\frac{1}{2}\lambda -\frac{1}{4}g^2+g'^2 \right) (\phi^\dagger \phi) \bar{\psi}_L \sqrt{2}H\mathcal{Y} \psi_R + \text{h.c.} \; , \\
-\frac{1}{3}\text{tr }[D^\mu,N_{\mu\nu}]b^\nu =& -\frac{5}{6}g'^2(\phi^\dagger D_\mu \phi)^*(\phi^\dagger D^\mu \phi) -\left( \frac{1}{8}g^2+\frac{5}{24}g'^2 \right) (\phi^\dagger \phi)\Box(\phi^\dagger \phi) \\ &+\frac{1}{6}g^2m^2(\phi^\dagger\phi)^2 -\frac{1}{6}\lambda g^2(\phi^\dagger\phi)^3 -\frac{1}{12}g^2(\phi^\dagger \phi)\bar{\psi}_L\sqrt{2}H\mathcal{Y}\psi_R  \\ &-\frac{5}{12}g'^2(\phi^\dagger i\overset{\text{\footnotesize{$\leftrightarrow$}}}{D_\mu}\phi) \bar{\psi}_R\gamma^\mu Y_R \psi_R -\frac{5}{12}g'^2(\phi^\dagger i\overset{\text{\footnotesize{$\leftrightarrow$}}}{D_\mu}\phi) \bar{\psi}_L \gamma^\mu Y_L \psi_L \\ &-\frac{1}{24}g^2(\phi^\dagger i\overset{\text{\footnotesize{$\leftrightarrow$}}}{D^a_\mu}\phi)\bar{\psi}_L \gamma^\mu \sigma^a \psi_L + \text{h.c.} \; ,\\
-\text{tr }aM^2 =& -2g^2(\phi^\dagger D_\mu \phi)^*(\phi^\dagger D^\mu \phi) -\left(-\frac{1}{2}g^2+\frac{1}{2}g'^2\right)(\phi^\dagger \phi)\Box(\phi^\dagger \phi) \\ &-\left(\lambda^2 -\lambda g^2 + \lambda g'^2 +\frac{1}{4}g^4 +\frac{1}{2}g^2 g'^2 +\frac{1}{4}g'^4 \right)(\phi^\dagger \phi)^3 \\ &+\left(2\lambda -g^2 +g'^2 \right) m^2 (\phi^\dagger \phi)^2 -m^4(\phi^\dagger \phi)\hs \\ &+g^2 (\phi^\dagger \phi) \bar{\psi}_L \sqrt{2}H\mathcal{Y} \psi_R + \text{h.c.} \; ,\hs \\ \text{tr }i\bar{\Gamma}\slashed{b} \Gamma =&\ \frac{1}{8}(\phi^\dagger i\overset{\text{\footnotesize{$\leftrightarrow$}}}{D^a_\mu}\phi)\bar{\psi}_L \gamma^\mu \sigma^a \langle \mathcal{Y}\mathcal{Y}^\dagger \rangle_I \psi_L +\frac{5}{2}(\phi^\dagger i\overset{\text{\footnotesize{$\leftrightarrow$}}}{D_\mu}\phi)\bar{\psi}_R \gamma^\mu \mathcal{Y}^\dagger \mathcal{Y} \tau^3 \psi_R \\ &-\frac{5}{4}(\phi^\dagger i\overset{\text{\footnotesize{$\leftrightarrow$}}}{D_\mu}\phi)\bar{\psi}_L\gamma^\mu \left(2\mathcal{Y}\mathcal{Y}^\dagger -\langle \mathcal{Y} \mathcal{Y}^\dagger \rangle_I \right)\tau^3 \psi_L \\ &-i\bar{\psi}_R\gamma^\mu \mathcal{Y}^\dagger \vp\left(\tau^1 t_R^1 + \tau^2 t_R^2\right)D_\mu\vp \mathcal{Y} \psi_R \hs   \\&-\left( \frac{1}{2}g^2+\frac{1}{2}g'^2 \right)(\phi^\dagger \phi) \bar{\psi}_L \sqrt{2}H\mathcal{Y} \psi_R +\text{h.c.} \; , \\ -\frac{1}{6}\text{tr }\bar{\Gamma} i\overset{\text{\footnotesize{$\leftrightarrow$}}}{\mathcal{D}}_{a} \Gamma  =&\ \frac{1}{2}(\phi^\dagger\phi) \bar{\psi}_L \sqrt{2}H\mathcal{Y}\mathcal{Y}^\dagger\mathcal{Y} \psi_R + \text{h.c.}  \label{phiR_contributions}
\end{aligned}
\end{align}

We reduce contractions with the Levi-Civita tensor using 
\begin{align}
\epsilon_{ijkl} (D_\mu\vp)_i (t_R^3D^\mu\vp)_j (t_R^3\vp)_k \vp_l = 
(\vp t_R^3 D_\mu \vp)^2 - \frac{1}{4}(\vp D_\mu \vp)^2 +
\frac{1}{4}\vp^2 (D_\mu \vp)^2 \; .
\end{align}
Combining the results in (\ref{box_div}) and (\ref{phiR_contributions})
according to (\ref{boxdr}), we obtain
\begin{align}
\begin{aligned}
32\pi^2\epsilon\mathcal{L}^\text{div}_{\phi D} = \frac{C_{\phi D}}{\Lambda^2} \bigg( &-\left(6\lambda+\frac{33}{2}g^2+\frac{19}{6}g'^2 \right)(\phi^\dagger D_\mu \phi)^*(\phi^\dagger D^\mu \phi) +m^2\bar{\psi}_L\sqrt{2}H\mathcal{Y} \psi_R  \\ &-\frac{5}{3}g'^2(\phi^\dagger \phi)\Box(\phi^\dagger \phi) +\left(12\lambda -3g^2+3g'^2 \right)m^2(\phi^\dagger\phi)^2 \\ &-\left(12\lambda^2-3\lambda g^2 +3\lambda g'^2 -\frac{3}{4}g^4 -\frac{3}{2}g^2g'^2 -\frac{3}{4}g'^4\right)(\phi^\dagger \phi)^3  \\&-(\phi^\dagger\phi)\bar{\psi}_L\sqrt{2}H\left(\lambda\mathcal{Y}-\frac{3}{2}g^2\mathcal{Y}+\frac{3}{2}g'^2\mathcal{Y} +\mathcal{Y}\mathcal{Y}^\dagger\mathcal{Y} \right)\psi_R \\ &-(\phi^\dagger i\overset{\text{\footnotesize{$\leftrightarrow$}}}{D_\mu}\phi)\bar{\psi}_R\gamma^\mu\left(\frac{1}{3}g'^2 Y_R -2\tau^3\mathcal{Y}^\dagger \mathcal{Y} \right)\psi_R -4m^4(\phi^\dagger \phi) \\&-(\phi^\dagger i\overset{\text{\footnotesize{$\leftrightarrow$}}}{D_\mu}\phi)\bar{\psi}_L\gamma^\mu\left(\frac{1}{3}g'^2Y_L +\left( 2\mathcal{Y}\mathcal{Y}^\dagger -\langle\mathcal{Y}\mathcal{Y}^\dagger\rangle_I\right)\tau^3 \right) \psi_L \\&-2i\bar{\psi}_R\gamma^\mu \mathcal{Y}^\dagger \vp(\tau^1 t_R^1 + \tau^2t_R^2)D_\mu\vp \mathcal{Y}\psi_R + \text{h.c.}\bigg) \; .
\end{aligned}
\end{align}
We find the following set of beta-function entries:
\begin{align}\begin{aligned}
\beta_{\phi D} &\supseteq \left(6\lambda +\frac{9}{2}g^2 -\frac{5}{6}g'^2 +4\gamma_\phi\right)C_{\phi D}\; , \qquad &\beta_{\phi\Box} &\supseteq \frac{5}{3}g'^2 C_{\phi D} \; ,
 \\ \beta_\phi &\supseteq \left(12\lambda^2 - 3 \lambda (g^2-g'^2) -\frac{3}{4}(g^2+g'^2)^2\right)C_{\phi D}\; , \qquad &\beta_{m^2} &\supseteq 4\frac{m^4}{\Lambda^2}C_{\phi D} \; , \\ \beta_\lambda &\supseteq\left(24\lambda -6(g^2-g'^2)\right)\frac{m^2}{\Lambda^2}C_{\phi D} \; , &\beta_{\mathcal{Y}} &\supseteq \frac{m^2}{\Lambda^2}\mathcal{Y}C_{\phi D} \; , \\ \beta_{\psi\phi}&\supseteq\left(\lambda\mathcal{Y} -\frac{3}{2}(g^2-g'^2)\mathcal{Y} +\mathcal{Y}\mathcal{Y}^\dagger\mathcal{Y} \right)C_{\phi D} \; , &\beta_{\phi ud} &\supseteq -\mathcal{Y}_u^\dagger \mathcal{Y}_d C_{\phi D}\; , \\ \beta_{\phi\psi} &\supseteq \left(\frac{1}{3}g'^2Y_R-2\tau^3\mathcal{Y}^\dagger\mathcal{Y}\right) C_{\phi D} \; , \\
 \beta_{\phi\psi}^{(1)} &\supseteq \left(\frac{1}{3}g'^2Y_L +\left( 2\mathcal{Y}\mathcal{Y}^\dagger -\langle\mathcal{Y}\mathcal{Y}^\dagger\rangle_I\right)\tau^3 \right)C_{\phi D} \; . \label{beta_phiD}
\end{aligned}
\end{align}

This completes our calculation for the bosonic operators.

\section{Summary of results} 
\label{sec:summary}

We summarize our results by listing all the contributions to the
renormalization-group beta functions that arise from the pure bosonic 
operators of dimension 6 in SMEFT. All results are in agreement with
\cite{Jenkins:2013zja,Jenkins:2013wua,Alonso:2013hga}
(see also the compilation in \cite{Celis:2017hod}). 

Adding the individual contributions derived in section \ref{sec:bosonic} 
we obtain:
\begin{align}
\beta_G  \supseteq 15g_s^2 C_G \; , \qquad 
\beta_{\wt{G}}  \supseteq 15g_s^2 C_{\wt{G}} \; , \qquad   
\beta_W  \supseteq \frac{29}{2} g^2 C_W \; , \qquad 
\beta_{\wt{W}}  \supseteq \frac{29}{2} g^2 C_{\wt{W}} 
\end{align}

\begin{align}
\beta_{\phi G} \supseteq \left(6\lambda -14g_s^2 - \frac{9}{2}g^2 
-\frac{3}{2}g'^2 +2\gamma_\phi \right) C_{\phi G} 
\end{align}

\begin{align}
\beta_{\phi\wt{G}} &\supseteq \left(6\lambda -14g_s^2 - \frac{9}{2}g^2 -
\frac{3}{2}g'^2 + 2\gamma_\phi \right) C_{\phi\wt{G}} 
\end{align}

\begin{align}
\beta_{\phi B} \supseteq \left(6\lambda -\frac{9}{2}g^2 + \frac{85}{6}g'^2 
+2\gamma_\phi \right) C_{\phi B} + 3gg' C_{\phi W B} 
\end{align}

\begin{align}
\beta_{\phi\wt{B}} \supseteq \left(6\lambda -\frac{9}{2}g^2 +\frac{85}{6}g'^2 + 
2\gamma_\phi \right) C_{\phi\wt{B}} + 3gg' C_{\phi \wt{W} B} \; , 
\end{align}

\begin{align}
\beta_{\phi W} \supseteq -15g^3C_W  + \left(6\lambda -\frac{53}{6}g^2 - 
\frac{3}{2}g'^2 + 2\gamma_\phi \right) C_{\phi W} + gg'C_{\phi W B} 
\end{align}

\begin{align}
\beta_{\phi \wt{W}} \supseteq -15g^3C_{\wt{W}} +\left(6\lambda -\frac{53}{6}g^2 
- \frac{3}{2}g'^2 + 2\gamma_\phi \right) C_{\phi\wt{W}} + gg' C_{\phi \wt{W} B}
\end{align}

\begin{align} 
\beta_{\phi W B} \supseteq 3g^2 g' C_W +2gg' C_{\phi W} +2gg' C_{\phi B} + 
\left(2\lambda +\frac{4}{3}g^2+\frac{19}{3}g'^2 +2\gamma_\phi \right)C_{\phi W B} 
\end{align}

\begin{align} 
\beta_{\phi \wt{W} B} \supseteq 3g^2 g' C_{\wt{W}} +2gg' C_{\phi\wt{W}}  
+2gg' C_{\phi\wt{B}}  +\left(2\lambda +\frac{4}{3}g^2 + 
\frac{19}{3}g'^2+2\gamma_\phi \right) C_{\phi \wt{W} B} 
\end{align}

\begin{align}
\beta_\phi &\supseteq \left(18\lambda g^2-9g^4 - 3g^2 g'^2 \right) C_{\phi W}
+\left(6\lambda g'^2 -3g^2g'^2 -3g'^4 \right) C_{\phi B} 
\nonumber\\
&+\left(6\lambda gg' -3g^3g' -3gg'^3 \right)C_{\phi W B} 
+\left(54\lambda  -\frac{27}{2}g^2 - \frac{9}{2}g'^2 + 
6\gamma_\phi \right)C_{\phi}
\nonumber\\
&+\left(-40\lambda^2+\frac{20}{3}\lambda g^2 \right) C_{\phi\Box} 
+\left(12\lambda^2 - 3 \lambda (g^2-g'^2) 
-\frac{3}{4}(g^2+g'^2)^2\right)C_{\phi D}
\end{align}

\begin{align}
\beta_{\phi\Box} \supseteq \left(12\lambda -4g^2-\frac{4}{3}g'^2 
+4\gamma_\phi \right)C_{\phi\Box} + \frac{5}{3}g'^2 C_{\phi D}
\end{align}

\begin{align}
\beta_{\phi D} \supseteq \frac{20}{3}g'^2 C_{\phi\Box} 
+\left(6\lambda +\frac{9}{2}g^2 -\frac{5}{6}g'^2 +4\gamma_\phi\right)C_{\phi D}
\end{align}

\begin{align}
\beta_{q G} \supseteq 9g_s^2C_G \mathcal{Y}_q +9ig_s^2 C_{\wt{G}}\mathcal{Y}_q
-4g_s\mathcal{Y}_q C_{\phi G} 
-4ig_s \mathcal{Y}_q C_{\phi\wt{G}}
\end{align}

\begin{align}
\beta_{\psi W} \supseteq -g\mathcal{Y} C_{\phi W}  -ig\mathcal{Y} C_{\phi\wt{W}} 
+2g'(Y_L+Y_R)\tau^3\mathcal{Y} C_{\phi W B}
+2ig'(Y_L+Y_R)\tau^3\mathcal{Y}C_{\phi \wt{W} B} 
\end{align}

\begin{align}
\beta_{\psi B} \supseteq -2g'(Y_L + Y_R) \mathcal{Y}  C_{\phi B}  
-2ig'(Y_L+Y_R) \mathcal{Y}  C_{\phi \wt{B}}  +3g\tau^3 \mathcal{Y} C_{\phi W B}  
+3ig\tau^3\mathcal{Y} C_{\phi \wt{W} B} 
\end{align}

\begin{align}
\beta_{q\phi} \supseteq 32g_s^2\mathcal{Y}_q C_{\phi G} 
+32ig_s^2 \mathcal{Y}_q C_{\phi\wt{G}}
\end{align}

\begin{align}
\beta_{\psi\phi} &\supseteq  9g^2 \mathcal{Y} C_{\phi W}    
+12g'^2(Y_L^2+Y_R^2)\mathcal{Y}C_{\phi B}
+9ig^2 \mathcal{Y} C_{\phi\wt{W}} 
\nonumber\\
&+12ig'^2 (Y_L^2+Y_R^2) \mathcal{Y}  C_{\phi \wt{B}}
-12gg'Y_L\tau^3\mathcal{Y}C_{\phi W B}   
-12igg'Y_L\tau^3 \mathcal{Y}C_{\phi \wt{W} B}
\nonumber\\ 
&+\left(-2\lambda \mathcal{Y} +\frac{10}{3}g^2\mathcal{Y} -6\mathcal{Y}
\mathcal{Y}^\dagger \mathcal{Y}\right) C_{\phi\Box} 
+\left(\lambda\mathcal{Y} -\frac{3}{2}(g^2-g'^2)\mathcal{Y} 
+\mathcal{Y}\mathcal{Y}^\dagger\mathcal{Y} \right)C_{\phi D} 
\end{align}

\begin{align}
\beta_{\phi\psi} \supseteq \left(\frac{1}{3}g'^2Y_R
-2\tau^3\mathcal{Y}^\dagger\mathcal{Y}\right) 
\left(C_{\phi\Box} + C_{\phi D}\right) \; , \qquad
\beta_{\phi ud} \supseteq \mathcal{Y}^\dagger_u \mathcal{Y}_d 
\left( 2 C_{\phi\Box} - C_{\phi D}\right)
\end{align}

\begin{align}
\beta^{(1)}_{\phi\psi} \supseteq \left(\frac{1}{3}g'^2Y_L 
+\left(2\mathcal{Y}\mathcal{Y}^\dagger 
-\langle\mathcal{Y}\mathcal{Y}^\dagger \rangle_I\right)\tau^3 \right)
\left(C_{\phi\Box} + C_{\phi D}\right)
\end{align}

\begin{align}
\beta^{(3)}_{\phi\psi} \supseteq \left(\frac{1}{6}g^2
-\frac{1}{2}\langle\mathcal{Y}\mathcal{Y}^\dagger \rangle_I \right) C_{\phi\Box} 
\end{align}
Here $\langle\ldots\rangle_I$ denotes a trace over isospin indices.

Finally, we collect the contributions from the bosonic dimension-6
operators to the beta functions of couplings in the SM at dimension 4:  

\begin{align}
\beta_{g_s} \supseteq - 8g_s \frac{m^2}{\Lambda^2} C_{\phi G} \; , \qquad 
\beta_{g} \supseteq -8g\frac{m^2}{\Lambda^2} C_{\phi W} \; ,\qquad
\beta_{g'} \supseteq -8g' \frac{m^2}{\Lambda^2} C_{\phi B} 
\end{align}

\begin{align} 
\beta_\lambda &\supseteq 36g^2 \frac{m^2}{\Lambda^2} C_{\phi W} 
+12g'^2\frac{m^2}{\Lambda^2} C_{\phi B} 
+12gg' \frac{m^2}{\Lambda^2} C_{\phi W B} 
+48\frac{m^2}{\Lambda^2} C_{\phi} \nonumber\\ 
&+\left(-64\lambda+\frac{40}{3}g^2 \right) \frac{m^2}{\Lambda^2} C_{\phi\Box} 
+\left(24\lambda -6(g^2-g'^2)\right) \frac{m^2}{\Lambda^2}C_{\phi D}  
\end{align}

\begin{align}
\beta_{m^2} \supseteq  4\frac{m^4}{\Lambda^2}
\left(-2 C_{\phi\Box} + C_{\phi D}\right) \; , \qquad
\beta_\mathcal{Y} \supseteq \frac{m^2}{\Lambda^2}\mathcal{Y}
\left(-2 C_{\phi\Box} + C_{\phi D}\right) 
\end{align}

\begin{align}
\beta_{\theta_s} \supseteq -\frac{128\pi^2}{g_s^2} \frac{m^2}{\Lambda^2} 
C_{\phi\wt{G}} \; ,\qquad
 \beta_\theta \supseteq -\frac{128\pi^2}{g^2}\frac{m^2}{\Lambda^2} 
C_{\phi\wt{W}} \; , \qquad
\beta_{\theta'} \supseteq -\frac{128\pi^2}{g'^2}\frac{m^2}{\Lambda^2}C_{\phi\wt{B}} 
\end{align}

\section{Conclusions} 
\label{sec:conc}

We have shown how functional methods provide an efficient way
to compute UV divergences to one loop in SMEFT.
Using the background-field method and a super-heat-kernel expansion, 
we derived a master formula for the one-loop divergences of EFTs that
generalizes a known formula, originally due to 't Hooft~\cite{tHooft:1973bhk}.
The generalization allows for the addition of a non-standard term of the form
$a^{\mu\nu}D_\mu D_\nu$ to the fluctuation operator $\Delta =D^\mu D_\mu + Y$,
treated to first order in the field-dependent quantity $a^{\mu\nu}$.

As an application of this master formula we computed the complete one-loop
divergences from insertions of the purely bosonic dimension-6 operators
in the Warsaw basis~\cite{Grzadkowski:2010es} of SMEFT.
We derived the corresponding RGEs, describing the RG mixing of the bosonic 
dimension-6 operators into any SMEFT operator of dimension 4 or 6. Our 
analysis serves as an independent confirmation of results previously obtained 
in the literature~\cite{Jenkins:2013zja,Jenkins:2013wua,Alonso:2013hga}.

We have also discussed how the RG beta-functions (anomalous dimensions)
for operator coefficients in SMEFT are related to the one-loop
divergences, demonstrating that this relation is governed by
chiral dimensions.
In future work, we plan to return to the renormalization of the remaining
dimension-6 operators in SMEFT using functional methods.

\section*{Acknowledgements}

The work of G.B. and A.C. has been supported by the DFG grant  BU 1391/2-1.  
The work of C.K. is supported by the Alexander von Humboldt-Foundation and by 
the Fermi Research Alliance, LLC under Contract No. DE- AC02-07CH11359 with 
the U.S. Department of Energy, Office of Science, 
Office of High Energy Physics. 

\appendix

\section{Details on the Higgs field representation}  
\label{sec:app2}

We express the Higgs field degrees of freedom as ($j\in\{0,1,2,3\}$)    
\begin{align}
H \equiv \frac{1}{\sqrt{2}}\left(\widetilde{\phi},\phi\right)  
=   i\tau^j\vp_j \; ,
\end{align}
with $\tau^{a}$ ($a\in\{1,2,3\}$) the generators of $\mathrm{SU(2)}$ and 
$\tau^0=-\frac{i}{2}\mathbf{1}$.     Under an electroweak gauge transformation 
$H \to g_L H g_R^{\dag}$, with $g_L \in  \mathrm{SU}(2)_L$ and $g_R$ belonging 
to the $\mathrm{U(1)}_Y$ subgroup of $\mathrm{SU}(2)_R$.   Since 
$\mathrm{SU}(2) \otimes \mathrm{SU}(2) $ is the universal covering group of 
$\mathrm{SO}(4)$, we can express the transformation properties of $\vp_i$ in 
terms of $\mathrm{SO}(4)$ generators.  The covariant derivative acting on the 
fields $\vp_i$ is given by
\begin{align}
(D_\mu\vp)_i=
\partial_\mu \vp_i +igW^a_\mu t^a_{Lij}\vp_j +ig'B_\mu t^3_{Rij}\vp_j \; ,
\end{align}
with the $\mathrm{SO}(4)$ generators
\begin{align}
\begin{aligned}
t_{Lij}^a &\equiv+2\, \text{tr }(\tau^i)^\dagger \tau^a \tau^j  = 
-\frac{i}{2} \left( \epsilon^{aij} + \delta^{ai} \delta^{0j} - 
\delta^{0i} \delta^{aj}\right) \; ,\\
t_{Rij}^a &\equiv -2\, \text{tr }(\tau^i)^\dagger \tau^j \tau^a =
-\frac{i}{2} \left( \epsilon^{aij} - \delta^{ai} \delta^{0j} + 
\delta^{0i} \delta^{aj}\right) \; .
\end{aligned}
\end{align}
Here $a,b\in\{1,2,3\}$ and $i,j,k,l\in\{0,1,2,3\}$. The antisymmetric tensor 
$\epsilon^{aij}$ is defined such that $\epsilon^{aij}=0$ if $i=0$ or $j=0$.  
In matrix form we can write
\begin{align}
\begin{aligned}
t_L^1 &= -\frac{i}{2}\begin{pmatrix}0&-1&0&0\\1&0&0&0\\0&0&0&1\\0&0&-1&0\\ \end{pmatrix}\; , \,\,\quad    t_L^2=-\frac{i}{2}\begin{pmatrix}0&0&-1&0\\0&0&0&-1\\1&0&0&0\\0&1&0&0\\ \end{pmatrix} \; , \,\,\quad   t_L^3 = -\frac{i}{2}\begin{pmatrix}0&0&0&-1\\0&0&1&0\\0&-1&0&0\\1&0&0&0\\ \end{pmatrix}  \,, \\  t_R^1 &= -\frac{i}{2}\begin{pmatrix}0&1&0&0\\-1&0&0&0\\0&0&0&1\\0&0&-1&0\\ \end{pmatrix}\; , \,\,\quad  t_R^2=-\frac{i}{2}\begin{pmatrix}0&0&1&0\\0&0&0&-1\\-1&0&0&0\\0&1&0&0\\ \end{pmatrix} \; , \,\,\quad  t_R^3 = -\frac{i}{2}\begin{pmatrix}0&0&0&1\\0&0&1&0\\0&-1&0&0\\-1&0&0&0\\ \end{pmatrix} \; .\end{aligned} 
\end{align}
These matrices fulfill the $\mathrm{SU}(2)$ algebra
\begin{align}
\begin{aligned}
\big[t_L^a,t_L^b\big]&=i\epsilon^{abc}t_L^c \; ,\\ \big\{t_L^a,t_L^b\big\} &= \frac{1}{2}\delta^{ab} \; ,\\ \text{tr }t_L^at_L^b &= \delta^{ab} \; ,\end{aligned}
&& \begin{aligned} \big[t_R^a,t_R^b\big]&=i\epsilon^{abc}t_R^c \; , \\ \big\{t_R^a,t_R^b\big\} &= \frac{1}{2}\delta^{ab}\; , \\ \text{tr }t_R^at_R^b &= \delta^{ab} \; , \label{su2_rel}
\end{aligned}
\end{align}
with $\big[t_L^a,t_R^b\big] = 0$ and $\text{tr } t_L^at^b_R = 0$. Some useful identities are 
\begin{align}
\begin{aligned}
 t_{Lik}^at_{Lkj}^b &= \frac{1}{4} \delta^{ab}\delta_{ij} +\frac{i}{2}\epsilon^{abc}t^c_{Lij}\; ,  \qquad t_{Lij}^a t_{Lkl}^a= \frac{1}{4}\left(\delta_{il}\delta_{jk} - \delta_{ik}\delta_{jl} + \epsilon_{ijkl} \right)\,,   \\
 t_{Rik}^at_{Rkj}^b &= \frac{1}{4} \delta^{ab}\delta_{ij}+\frac{i}{2}\epsilon^{abc}t^c_{Rij} \,, \qquad t_{Rij}^a t_{Rkl}^a= \frac{1}{4}\left(\delta_{il}\delta_{jk} - \delta_{ik}\delta_{jl} - \epsilon_{ijkl} \right)\; .
\end{aligned}
\end{align}
Here $\epsilon_{ijkl}$ denotes the totally anti-symmetric 4-dimensional tensor.     Using the real representation for the Higgs field, the SM equations of motion read
\begin{align}
\begin{aligned}
(D_\mu G^{\mu\nu})^A &= g_s\bar{q}\gamma^\nu T^A q \vphantom{\frac{1}{2}}  \,, \\
(D_\mu W^{\mu\nu})^a &= igt_{Lij}^a \vp_i (D^\nu\vp)_j + g \bar{\psi}\gamma^\nu \tau^{a} P_L \psi \vphantom{\frac{1}{2}} \,, \\
\partial_\mu B^{\mu\nu} &= ig't_{Rij}^3 \vp_i (D^\nu\vp)_j +g'\bar{\psi}\gamma^\nu(Y_L P_L + Y_R P_R)\psi \vphantom{\frac{1}{2}} \,, \\
(D_\mu D^\mu \vp)_i &= m^2\vp_i -\frac{\lambda}{2}(\vp_j\vp_j)\vp_i - i\bar{\psi}\sqrt{2}\left(\tau^i\mathcal{Y}P_R-\mathcal{Y}^\dagger(\tau^i)^\dagger P_L\right)\psi\vphantom{\frac{1}{2}} \,, \\
i\slashed{D}\psi &= \sqrt{2}(H\mathcal{Y}P_R + \mathcal{Y}^\dagger H^\dagger P_L)\psi \vphantom{\frac{1}{2}} \,.
\end{aligned}
\end{align}

\section{SM fluctuation operator}   
\label{sec:appSM}

The SM fluctuation operator can be cast in the form of~\eqref{minform} by 
choosing the Feynman gauge.        In the electroweak sector we use a gauge 
fixing term that cancels the mixing between the gauge fields and the would-be 
Goldstone bosons in the SM~\cite{Denner:1994xt}.    For QCD we take the usual 
Yang-Mills gauge fixing~\cite{Abbott:1981ke}.     
The gauge-fixing Lagrangian reads\footnote{A general discussion
of gauge fixing in SMEFT can be found in \cite{Helset:2018fgq}}
\begin{align}
\mathcal{L}_\text{g.f.} = 
-\frac{1}{2} f^A f^A -\frac{1}{2} f^a f^a -\frac{1}{2} f^2 \label{gf_SM} \; .
\end{align}
with
\begin{align}
\begin{aligned}
f^A &= D_\mu \alpha^{A\mu} \;, \qquad 
f^a = D_\mu \omega^{a\mu} -igt^a_{Lij}\xi_i \vp_j \; ,\qquad 
f = \partial_\mu \beta^\mu -ig't^3_{Rij} \xi_i \vp_j \; . \label{SM_gauge_func}
\end{aligned}
\end{align}
The covariant derivatives act on $\alpha_\mu$ and $\omega_\mu$ in the adjoint 
representation.  The Faddeev--Popov Lagrangian is quadratic in the ghost 
fields, which do not mix with the other degrees of freedom, so its divergences 
can be calculated separately:
\begin{align}
32\pi^2\epsilon &\mathcal{L}_\text{div, \text{ghost}}^1 = \frac{1}{6}g_s^2C^\text{ad}_3G^A_{\mu\nu}G^{A\mu\nu} + \frac{1}{6}g^2C^\text{ad}_2W^a_{\mu\nu}W^{a\mu\nu} -\frac{\lambda}{2}(\phi^\dagger\phi)^2\left(\frac{3g^4+2g^2g'^2+g'^4}{2\lambda}\right) \; .
\end{align}
Here $C^\text{ad}_N = N$ is the Dynkin index of $\mathrm{SU(N)}$
in the adjoint representation.   The bosonic building blocks $N^{\mu}$ and $M$ in~\eqref{minform2} are given by
\begin{align}
\begin{aligned}
N_{IJ}^\mu &= \begin{pmatrix} g_s f^{ABC} G^{C\mu}g_{\alpha\beta} &&& \\ & g\epsilon^{abc}W^{c\mu}g_{\lambda\rho} &&\\&&0&\\&&& igW^{d\mu}t^d_{Lij} +ig'B^\mu t^3_{Rij} \end{pmatrix} \; , \\
M_{IJ} & = \begin{pmatrix}2g_sf^{ABC}G^C_{\alpha\beta}&0&0&0\\0&2g\epsilon^{abc}W^c_{\lambda\rho} + \dfrac{g^2}{4}(\vp_k\vp_k)\delta^{ab}g_{\lambda\rho}&gg'(\vp t_L^a t_R^3\vp) g_{\lambda\kappa}& -2g(t^a_L D_\lambda \vp)_j\\ 0&gg'(\vp t_L^b t_R^3\vp) g_{\rho\sigma}&\dfrac{g'^2}{4}(\vp_k\vp_k) g_{\sigma\kappa}& -2g'(t_R^3 D_\sigma \vp)_j\\0&-2g(t_L^b D_\rho \vp)_i &-2g'(t_R^3 D_\kappa \vp)_i & M_{ij}\end{pmatrix} \,,
\end{aligned}
\end{align}
with the field indices $I=(A\alpha,a\lambda,\sigma,i)$ and $J=(B\beta,b\rho,\kappa,j)$.  Here we have defined 
\begin{align}
M_{ij}\equiv\left( \left(\frac{\lambda}{2}+\frac{g^2}{4}\right)(\vp_k\vp_k)-m^2 \right)\delta_{ij} +\left(\lambda-\frac{g^2}{4}\right)\vp_i\vp_j-g'^2(t^3_R\vp)_i(t^3_R\vp)_j \; . \label{Mij}
\end{align}
The fermion-boson mixing terms in \eqref{eqtrick} are given by
\begin{align}
\Gamma^T &=(-i)\begin{pmatrix}g_sT^B\gamma^\beta\psi\\ g\tau^b\gamma^\rho P_L\psi\\g'\gamma^\kappa(Y_LP_L+Y_RP_R)\psi\\\sqrt{2}(\tau^j\mathcal{Y} P_R - \mathcal{Y}^\dagger (\tau^j)^\dagger P_L)\psi \end{pmatrix} \;, \quad \bar{\Gamma}= (-i)\begin{pmatrix}g_s\bar{\psi}T^A\gamma^\alpha\\ g\bar{\psi}\tau^a\gamma^\lambda P_L\\g'\bar{\psi}\gamma^\sigma(Y_LP_L+Y_RP_R)\\ \sqrt{2}\bar{\psi}(\tau^i\mathcal{Y} P_R - \mathcal{Y}^\dagger (\tau^i)^\dagger P_L)\end{pmatrix} \nonumber \;,  \\
\label{SM_blocks1}
\end{align}
Note that in our conventions $\bar\Gamma_I  = S_{IJ}\, \Gamma_J^{\dag} \gamma_0$ 
with $S= \mathrm{diag}(-1,-1,-1,1)$ and $I,J$ labels for the 
bosonic variables.   
Finally, the pure fermionic terms in~\eqref{Fterm} read
\begin{align}
r&=\sqrt{2}H\mathcal{Y}\;, \quad    R_\mu= g_s G_\mu + g'B_\mu Y_R \;,\nonumber 
\vphantom{\frac{1}{2}}\\
l&=\sqrt{2}\mathcal{Y}^\dagger H^\dagger \; , \quad   
L_\mu= g_s G_\mu +gW_\mu + g'B_\mu Y_L \;.  \vphantom{\frac{1}{2}}     
\label{SM_blocks2}
\end{align}
$Y_{L,R}$ are the hypercharge matrices
\begin{align}
Y_L&=\text{diag}\left(\frac{1}{6},\frac{1}{6},-\frac{1}{2},-\frac{1}{2}\right) 
\qquad \text{and} \qquad  
Y_R=\text{diag}\left(\frac{2}{3},-\frac{1}{3},0,-1\right)\; . 
\end{align}

With these building blocks, 
and using the master formula in (\ref{master_formula}),
we may verify the one-loop renormalization group equations of the SM. First, 
including the ghost contributions, we find for the one-loop divergences 
\begin{eqnarray}
&& 32\pi^2\epsilon\, {\cal L}^{\rm SM}_{\rm div} =\nonumber\\
&& -\frac{1}{2}\langle G^{\mu\nu} G_{\mu\nu}\rangle
\left( -\frac{22 N_c - 4 N_f}{3}\right) g^2_s 
+2 C_F g^2_s  \bar q i\!\not\!\! D q 
- 8 C_F g^2_s \left(\bar q \sqrt{2} H {\cal Y}P_R q + {\rm h.c.}\right)
\nonumber\\
&& -\frac{1}{2}\langle W^{\mu\nu}W_{\mu\nu}\rangle
\left(-\frac{44}{3}+\frac{2}{3}(N_c+1)f+\frac{1}{3}\right)g^2
-\frac{1}{4}B^{\mu\nu}B_{\mu\nu}
\left(\left(\frac{22 N_c}{27}+2\right)f+\frac{1}{3}\right)g'^2
\nonumber\\
&& +D^\mu\phi^\dagger D_\mu\phi\left(-6g^2-2 g'^2 
+2 \langle {\cal Y}^\dagger {\cal Y}\rangle\right)
+ m^2 \phi^\dagger\phi\left(-\frac{3}{2}g^2-\frac{1}{2}g'^2-6\lambda\right)
\nonumber\\
&&-\frac{\lambda}{2}(\phi^\dagger\phi)^2
\left(-3g^2-g'^2-12\lambda-\frac{3}{4\lambda}(3 g^4+2 g^2 g'^2 + g'^4)
+\frac{4}{\lambda}\langle ({\cal Y}^\dagger {\cal Y})^2\rangle\right)\nonumber\\
&& + \bar\psi_L\left(\frac{3}{2}g^2+2g'^2 Y^2_L\right)i\!\not\!\! D \psi_L
+\bar\psi_R\, 2g'^2 Y^2_R i\!\not\!\! D \psi_R
-8 g'^2\left(\bar\psi_L \sqrt{2} H Y_L{\cal Y} Y_R \psi_R + {\rm h.c.}\right)
\nonumber\\
&&+\bar\psi_L\langle{\cal Y}{\cal Y}^\dagger\rangle_I\, i\!\not\!\! D \psi_L
+2\bar\psi_R{\cal Y}^\dagger{\cal Y}\, i\!\not\!\! D \psi_R
-2\left(\bar\psi_L \sqrt{2}H (\langle{\cal Y}{\cal Y}^\dagger\rangle_I -
{\cal Y}{\cal Y}^\dagger){\cal Y}\psi_R +{\rm h.c.}\right)
\label{lsmdiv}
\end{eqnarray}
Here $\langle\ldots\rangle_I$ represents a trace over isospin indices only. 
$N_c=3$, $f=3$, and $N_f=6$ denote the number of colours,
fermion generations, and quark flavours, respectively, and 
$C_F=(N^2_c-1)/2N_c$.
The quark fields are written as $q=(u,d,0,0)^T$. 

From the divergences in (\ref{lsmdiv}) we obtain the beta functions of the SM: 
\begin{eqnarray}
\beta_{g_s} &=& -\frac{11 N_c - 2 N_f}{3} g^3_s = -7 g^3_s\\
\beta_g &=& -\left(\frac{22}{3}-\frac{N_c+1}{3}f-\frac{1}{6}\right)g^3
=-\frac{19}{6}g^3\\
\beta_{g'} &=& \left(\left(\frac{11 N_c}{27}+1\right)f+\frac{1}{6}\right)g'^3
=\frac{41}{6}g'^3\\
\beta_\lambda &=& -3(3g^2+g'^2)\lambda+12\lambda^2 +
\frac{3}{4}(3 g^4+2 g^2 g'^2 + g'^4)
+4 \lambda \langle {\cal Y}^\dagger {\cal Y}\rangle 
-4 \langle ({\cal Y}^\dagger {\cal Y})^2\rangle \\
\beta_{m^2} &=& m^2\left(-\frac{9}{2}g^2-\frac{3}{2}g'^2 + 6\lambda + 
2  \langle {\cal Y}^\dagger {\cal Y}\rangle\right)\\
\beta_{\cal Y} &=& \frac{3}{2}\left(2{\cal Y}{\cal Y}^\dagger  
-\langle{\cal Y}{\cal Y}^\dagger\rangle_I \right){\cal Y}\nonumber\\
&&-\left(\frac{9}{4}g^2 +\left(\frac{3}{4}+6 Y_L Y_R\right)g'^2
- \langle {\cal Y}^\dagger {\cal Y}\rangle 
+6 C_F g^2_s P_q\right){\cal Y} \label{betasm}
\end{eqnarray}
with $3/4+6 Y_L Y_R={\rm diag}(17/12,5/12,3/4,15/4)$, 
$P_q={\rm diag}(1,1,0,0)$, 
in agreement with the results compiled in \cite{Celis:2017hod}
(see also \cite{Buchalla:2017jlu} for further details).



\begin{thebibliography}{99}

\bibitem{Jack:1984vj}
  I.~Jack and H.~Osborn,
  Nucl.\ Phys.\ B {\bf 249} (1985) 472.
  
    
\bibitem{Lee:1984ud}
  C.~Lee and C.~Rim,
  Nucl.\ Phys.\ B {\bf 255} (1985) 439.
  
  
\bibitem{Neufeld:1998js}
  H.~Neufeld, J.~Gasser and G.~Ecker,
  Phys.\ Lett.\ B {\bf 438} (1998) 106
  [hep-ph/9806436].
  
\bibitem{Abbott:1981ke}
  L.~F.~Abbott,
  Acta Phys.\ Polon.\ B {\bf 13} (1982) 33.
    
  
\bibitem{Neufeld:1998mb}
  H.~Neufeld,
  Eur.\ Phys.\ J.\ C {\bf 7} (1999) 355
  [hep-ph/9807425].
  


\bibitem{Knecht:1999ag}
  M.~Knecht, H.~Neufeld, H.~Rupertsberger and P.~Talavera,
  Eur.\ Phys.\ J.\ C {\bf 12} (2000) 469
  [hep-ph/9909284].
  
  
\bibitem{Buchalla:2017jlu}
  G.~Buchalla, O.~Cat\`a, A.~Celis, M.~Knecht and C.~Krause,
  Nucl.\ Phys.\ B {\bf 928} (2018) 93
  [arXiv:1710.06412 [hep-ph]].
  
  
  
\bibitem{tHooft:1973bhk}
  G.~'t Hooft,
  Nucl.\ Phys.\ B {\bf 62} (1973) 444.
  
\bibitem{Agadjanov:2013lra}
  A.~Agadjanov, D.~Agadjanov, A.~Khelashvili and A.~Rusetsky,
  Eur.\ Phys.\ J.\ A {\bf 49} (2013) 120
  [arXiv:1307.1451 [hep-ph]].
    
  
\bibitem{Buchmuller:1985jz} 
  W.~Buchm\"uller and D.~Wyler,
  Nucl.\ Phys.\ B {\bf 268}, 621 (1986).

\bibitem{Grzadkowski:2010es} 
  B.~Grzadkowski, M.~Iskrzynski, M.~Misiak and J.~Rosiek,
  JHEP {\bf 1010}, 085 (2010)
  [arXiv:1008.4884 [hep-ph]].
  
 

\bibitem{Jenkins:2013zja}
  E.~E.~Jenkins, A.~V.~Manohar and M.~Trott,
  JHEP {\bf 1310} (2013) 087
  [arXiv:1308.2627 [hep-ph]].
 
\bibitem{Jenkins:2013wua}
  E.~E.~Jenkins, A.~V.~Manohar and M.~Trott,
  JHEP {\bf 1401} (2014) 035
  [arXiv:1310.4838 [hep-ph]].
  
\bibitem{Alonso:2013hga}
  R.~Alonso, E.~E.~Jenkins, A.~V.~Manohar and M.~Trott,
  JHEP {\bf 1404} (2014) 159
  [arXiv:1312.2014 [hep-ph]].
  
\bibitem{Alonso:2014zka}
  R.~Alonso, H.~M.~Chang, E.~E.~Jenkins, A.~V.~Manohar and B.~Shotwell,
  Phys.\ Lett.\ B {\bf 734} (2014) 302
  [arXiv:1405.0486 [hep-ph]].
  
  
  \bibitem{Janmaster}  
J.-N.~Toelstede, 
Master thesis Ludwig-Maximilians-Universit\"at M\"unchen (2018).  

  
\bibitem{Grojean:2013kd}
  C.~Grojean, E.~E.~Jenkins, A.~V.~Manohar and M.~Trott,
  JHEP {\bf 1304} (2013) 016
  [arXiv:1301.2588 [hep-ph]];
  J.~Elias-Miro, J.~R.~Espinosa, E.~Masso and A.~Pomarol,
  JHEP {\bf 1311} (2013) 066
  [arXiv:1308.1879 [hep-ph]];
  JHEP {\bf 1308} (2013) 033
  [arXiv:1302.5661 [hep-ph]];
  G.~M.~Pruna and A.~Signer,
  JHEP {\bf 1410} (2014) 014
  [arXiv:1408.3565 [hep-ph]];
  V.~Cirigliano, W.~Dekens, J.~de Vries and E.~Mereghetti,
  Phys.\ Rev.\ D {\bf 94} (2016) no.3,  034031
  [arXiv:1605.04311 [hep-ph]];
  F.~Feruglio, P.~Paradisi and A.~Pattori,
  Phys.\ Rev.\ Lett.\  {\bf 118} (2017) no.1,  011801
  [arXiv:1606.00524 [hep-ph]];
  C.~Bobeth, A.~J.~Buras, A.~Celis and M.~Jung,
  JHEP {\bf 1704} (2017) 079
  [arXiv:1609.04783 [hep-ph]];
  A.~Celis, J.~Fuentes-Martin, A.~Vicente and J.~Virto,
  Phys.\ Rev.\ D {\bf 96} (2017) no.3,  035026
  [arXiv:1704.05672 [hep-ph]];
  C.~Bobeth, A.~J.~Buras, A.~Celis and M.~Jung,
  JHEP {\bf 1707} (2017) 124
  [arXiv:1703.04753 [hep-ph]];
  C.~Bobeth and A.~J.~Buras,
  JHEP {\bf 1802} (2018) 101
  [arXiv:1712.01295 [hep-ph]];
  A.~J.~Buras and M.~Jung,
  JHEP {\bf 1806} (2018) 067
  [arXiv:1804.05852 [hep-ph]];
  J.~Aebischer, C.~Bobeth, A.~J.~Buras and D.~M.~Straub,
  arXiv:1808.00466 [hep-ph];
  J.~E.~Camargo-Molina, A.~Celis and D.~A.~Faroughy,
  arXiv:1805.04917 [hep-ph];
  J.~Aebischer, J.~Kumar and D.~M.~Straub,
  arXiv:1804.05033 [hep-ph];
  M.~Chala, C.~Krause and G.~Nardini,
  JHEP {\bf 1807} (2018) 062
  [arXiv:1802.02168 [hep-ph]].
    
  
\bibitem{Celis:2017hod}
  A.~Celis, J.~Fuentes-Martin, A.~Vicente and J.~Virto,
  Eur.\ Phys.\ J.\ C {\bf 77} (2017) no.6,  405
  [arXiv:1704.04504 [hep-ph]].
  
  
\bibitem{VanNieuwenhuizen:1981ae}
  P.~Van Nieuwenhuizen,
  Phys.\ Rept.\  {\bf 68} (1981) 189.
  
\bibitem{Donoghue:1992dd}
  J.~F.~Donoghue, E.~Golowich and B.~R.~Holstein,
  {\it Dynamics of the standard model},
  Camb.\ Monogr.\ Part.\ Phys.\ Nucl.\ Phys.\ Cosmol.\  {\bf 35} (2014).
     
\bibitem{Ball:1988xg}
  R.~D.~Ball,
  Phys.\ Rept.\  {\bf 182} (1989) 1.
  
\bibitem{err}
  \href{http://einstein.ucsd.edu/smeft/}{http://einstein.ucsd.edu/smeft/}
  
\bibitem{Jenkins:2013sda}
  E.~E.~Jenkins, A.~V.~Manohar and M.~Trott,
  Phys.\ Lett.\ B {\bf 726} (2013) 697
  [arXiv:1309.0819 [hep-ph]].

\bibitem{Buchalla:2013eza}
  G.~Buchalla, O.~Cat\`a and C.~Krause,
  Phys.\ Lett.\ B {\bf 731} (2014) 80
  [arXiv:1312.5624 [hep-ph]].

\bibitem{Mertig:1990an}
  R.~Mertig, M.~Bohm and A.~Denner,
  Comput.\ Phys.\ Commun.\  {\bf 64} (1991) 345.

\bibitem{Shtabovenko:2016sxi}
  V.~Shtabovenko, R.~Mertig and F.~Orellana,
  Comput.\ Phys.\ Commun.\  {\bf 207} (2016) 432
  [arXiv:1601.01167 [hep-ph]].

\bibitem{wolframresearch}
  Wolfram Research, Inc., “Mathematica”, Version 11.1 (2017).

\bibitem{Borodulin:2017pwh}
  V.~I.~Borodulin, R.~N.~Rogalyov and S.~R.~Slabospitskii,
  arXiv:1702.08246 [hep-ph].

\bibitem{Denner:1994xt}
  A.~Denner, G.~Weiglein and S.~Dittmaier,
  Nucl.\ Phys.\ B {\bf 440} (1995) 95
  [hep-ph/9410338].

\bibitem{Helset:2018fgq}
  A.~Helset, M.~Paraskevas and M.~Trott,
  Phys.\ Rev.\ Lett.\  {\bf 120} (2018) no.25,  251801
  [arXiv:1803.08001 [hep-ph]].


\end{thebibliography}
\end{document}